# JingZhao: A Framework for Rapid NIC Prototyping in the Domain-Specific-Network Era

Fan Yang, yangfan2020@ict.ac.cn, Institute of Computing Technology, Chinese Academy of Sciences, China
Zhan Wang, wangzhan@ncic.ac.cn, Institute of Computing Technology, Chinese Academy of Sciences, China
Ning Kang, kangning@ict.ac.cn, Institute of Computing Technology, Chinese Academy of Sciences, China
Zhenlong Ma, mazhenlong@ncic.ac.cn, Institute of Computing Technology, Chinese Academy of Sciences, China
Jianxiong Li, lijianxiong20g@ict.ac.cn, Institute of Computing Technology, Chinese Academy of Sciences, China
Guojun Yuan, yuanguojun@ict.ac.cn, Institute of Computing Technology, Chinese Academy of Sciences, China
Guangming Tan, tgm@ncic.ac.cn, Institute of Computing Technology, Chinese Academy of Sciences, China

**Abstract**

The network is becoming domain-specific, which requires on-demand design of the network protocols, as well as the microarchitecture of the NIC. However, to develop such a NIC is not that easy. Since the scissor gap between network speed and the growth of CPU frequency is expanding, most of the protocols need to be offloaded to hardware. The process of designing, verifying and optimizing a domain-specific NIC usually takes great effort, which hinders the rapid iteration of new protocols and algorithms.

In this paper, we propose JingZhao, an open-sourced framework for NIC prototyping, which could be leveraged to rapidly implement and verify a domain-specific NIC. Using this framework, we implement a fully-functional RDMA NIC (RNIC). To the best of our knowledge, this represents the first open-source RDMA NIC solution with complete compatibility to the standard OFED communication library. The RNIC was also taped out using TSMC's 28nm process to validate our design. Our evaluation results show that new network functions can be easily integrated into the framework, and achieve nearly line-rate packet processing.

## 1 Introduction

*The era of Domain-Specific-Network is coming.*

Over the past decade, we have witnessed that as Moore's Law continues to slow down, the growing disparity between network speeds and CPU (Central Processing Unit) frequency scaling has led to progressively compressed instruction windows for packet processing[1]. As illustrated in **Figure 1**, sustaining high-speed network links now demands an ever-increasing number of CPU cores. Concurrently, we are seeing explosive growth in system scale (for instance, modern LLM training clusters now comprise over 10,000 nodes[2][3][4]), where the communication performance between GPUs (Graphics Processing Unit) or NPUs (Neural Processing Unit) has become the critical factor determining system scalability. To better overlap communication with computation, offloading the network stack to NIC (Network Interface Card) hardware has become imperative - their highly parallelized, dataflow-based architectures are inherently optimal for protocol processing. The most prominent example of such offloading is RDMA (Remote Direct Memory Access)[5], a technology that has been widely deployed in HPC (High Performance Computing) systems for decades and is now gaining widespread adoption in data centers due to its capabilities of ultra-low latency and high-bandwidth.

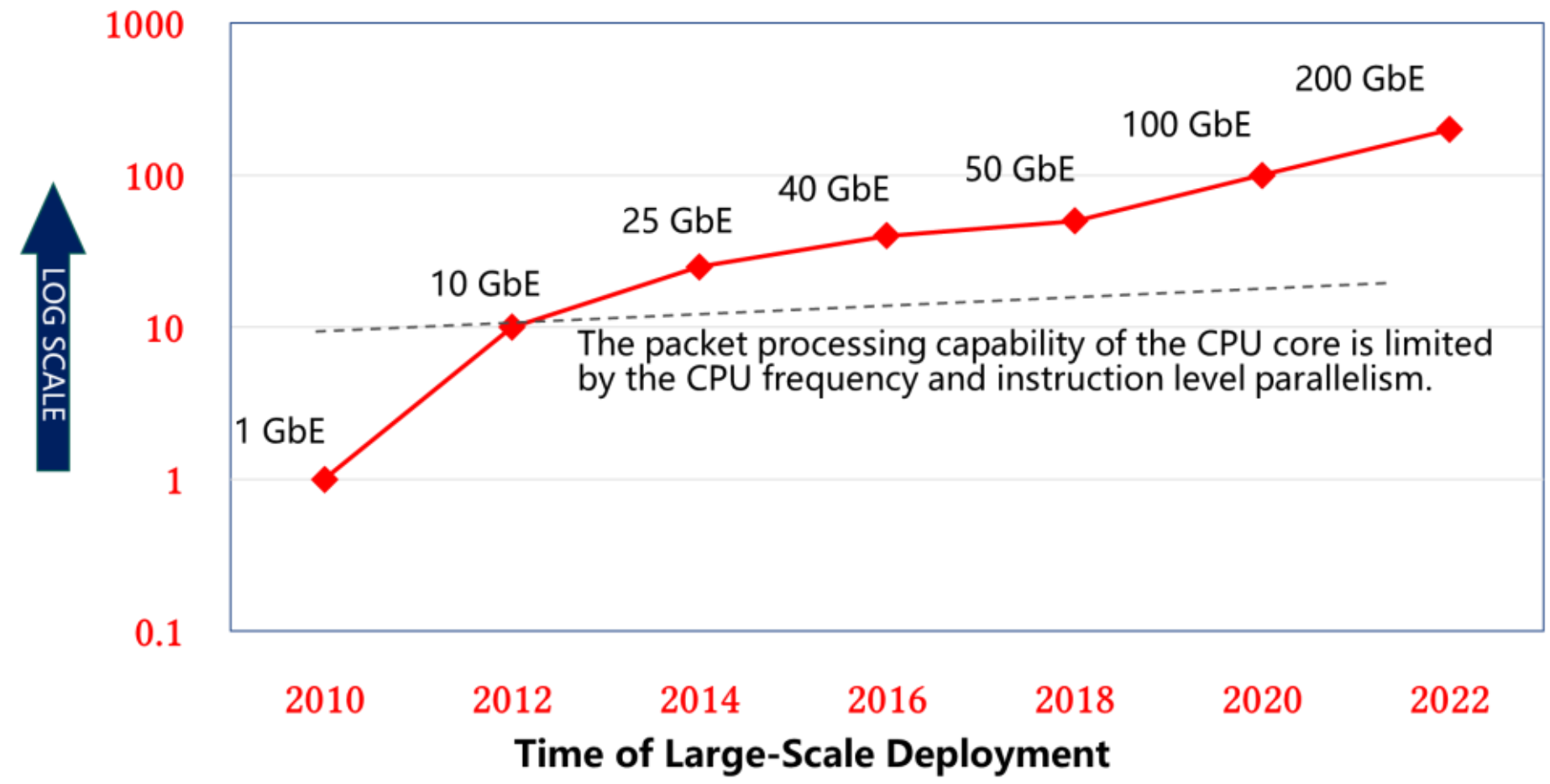

**Figure 1**: **Scissor gap between network speed and the growth of CPU frequency**

However, with the emergence of new application scenarios, the characteristics of communication behaviors exhibit significant differences, making traditional "one-size-fits-all" network designs inadequate to meet these demands. Taking

LLM training as an example: its traffic pattern is distinctly characterized by low entropy, high predictability, and sensitivity to tail latency—features that traditional RDMA networks struggle to accommodate[2]. In response, the industry has seen the rise of numerous domain-specific network protocols designed to enhance communication efficiency, including solutions for congestion control [6][7][8][9], load balancing[10][11], and packet loss recovery [12][13][14][15].

Based on these technical trends, we argue that networks are evolving toward increasing domain specialization, necessitating on-demand design of both network protocols and NIC microarchitectures. However, hardware offloading of new protocols to NICs presents significantly greater challenges than CPU implementation: developing new protocols requires substantial hardware engineering efforts, including managing low-level details such as inter-register/SRAM signal transmission; while verification and performance tuning demand meticulous optimization of interactions between the NIC, system buses (e.g., PCIe), and memory subsystems—all of which severely impede rapid iteration of new protocols and algorithms. This leads to our fundamental research question: how can we accelerate this process with reduced engineering effort? In response, we present JingZhao[1], an open-source framework for rapid NIC prototyping that enables efficient hardware offloading and fast validation of network protocols. We make the following concrete contributions:

1) We have proposed a high-performance NIC "template" and provide multiple high-performance, reusable building blocks that enable rapid construction of protocol processing pipelines, significantly reducing the development complexity and effort required for hardware protocol offloading;

2) We have designed a system-level simulation framework capable of precisely modeling interactions between high-performance NICs, system buses, and NIC drivers in real systems, facilitating bandwidth/latency/throughput evaluation and subsequent performance tuning;

3) Using this framework, we have implemented a fully-functional RDMA NIC (RNIC) as a concrete case study. To our knowledge, this is the first open-source RNIC with full compatibility to standard OFED communication library. The RNIC was also taped out using TSMC's 28nm process to validate our design;

4) We conducted comprehensive micro-benchmarking and system-level evaluations of both the framework and our RNIC, demonstrating their capability to achieve high-speed packet processing.

## 2 Background and Motivation

This section first reviews the working model and design challenges of modern high-performance NICs, and then discuss the requirements of rapid NIC prototyping.

### 2.1 Functional Model of a High-Performance NIC

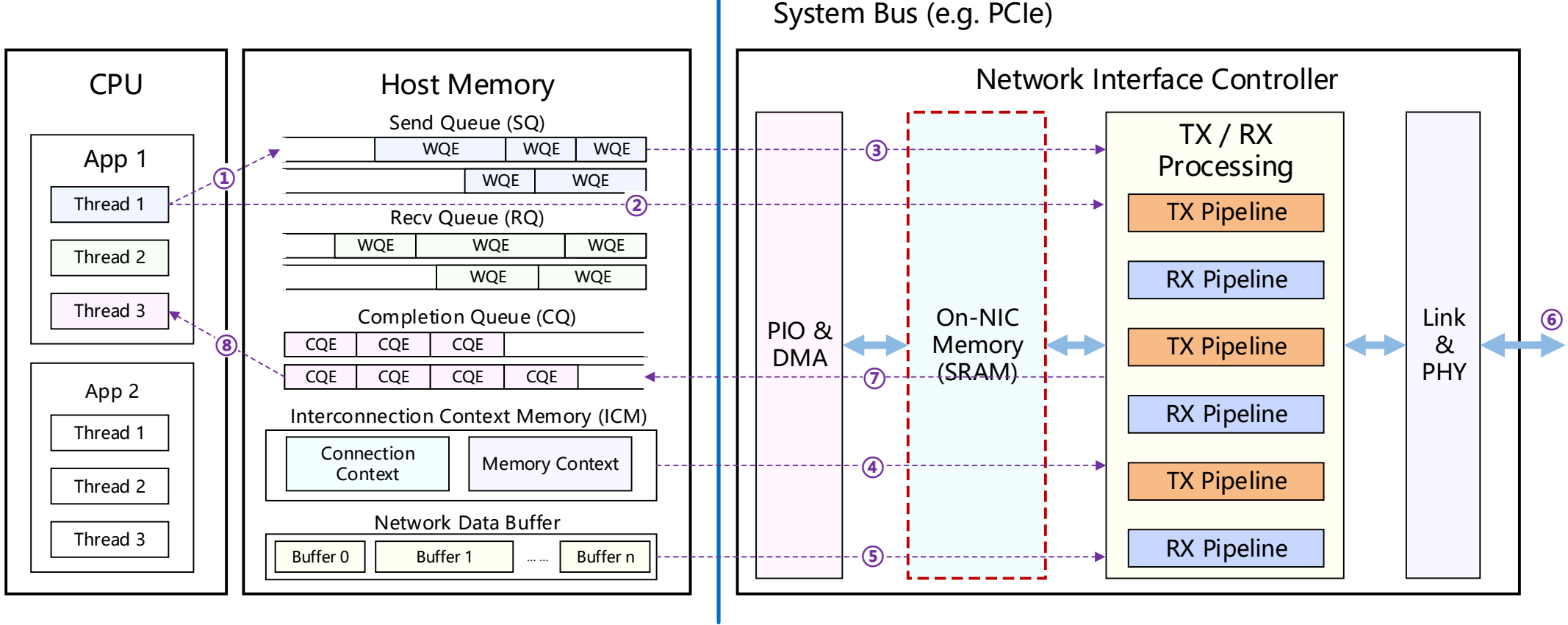

**Figure 2: Software-Hardware Interaction Model of a Modern NIC**

Modern NICs usually provide applications with asynchronous queue-based interfaces, which helps to overlap the processes of computation and communication. To better understand the interaction between application and NIC, let's take a simplified workflow of sending data as an example. As shown in **Figure 1**, when an application wants to send data, it first registers a data buffer, and then generates a Work Queue Element (WQE), which describes the memory address of the data buffer. The WQE is enqueued into the Send Queue (SQ). The above process is usually called Post Send (①). Subsequently, a register on the NIC is updated through Programmed I/O (PIO) to notify the NIC there's a new request to be handled. This process is also called Ring Doorbell (②). After completing these operations, the application can continue its computation task, while the remaining communication task is all taken over by the NIC. After receiving the doorbell, the NIC fetches the WQE from the SQ through Direct Memory Access (DMA) (③). At the same time, it also fetches the Context information from the memory (④), including Connection Context (e.g. source/destination IP/MAC address, packet sequence number, etc.), and Memory Context (e.g. metadata of the data buffer, protection domain, etc.). This information is stored in a separate area called Interconnection Context Memory (ICM), and will be further used in TX/RX pipeline processing. According to the memory address carried in WQE, the NIC DMA Reads data from host memory (⑤), generates a packet header according to different network protocols, and then sends the packet to the network (⑥). After that, the NIC generates a Completion Queue Event (CQE) and DMA Writes it to the Completion Queue (CQ) (⑦). The CQE is a notifier indicating the status of WQE processing. By polling the CQ, the application will learn whether the WQE is successfully processed or not (⑧). The Receive process is similar and will not be further discussed.

[1] https://github.com/ETH-PLUS/HanGuHTN

In the above process, the TX/RX pipelines need to interact with the host memory through system bus (typically PCIe) many times, which brings significant latency[5]. For example, the end-to-end latency of Mellanox CX-6 NIC is about 1.5µs[16], while a single PCIe traverse contributes about 300ns. To improve performance, the NIC usually maintains an on-chip SRAM to cache part of the host memory (including SQs, RQs and ICM). Depending on the requirements of different network protocols, SRAM may be centralized (all pipelines share a large block of SRAM, typically used for shared data structures, such as flow states) or distributed (different pipeline stages own their SRAMs exclusively, typically for storing independent packet header fields). When handling stateful protocols, the NIC needs to look up or modify communication states stored in the internal cache. Once a cache miss occurs, it must fetch the state information from host memory via PCIe and update its internal cache, resulting in additional overhead and performance degradation[16][17][18][19].

### 2.2 Problem Statement

Designing and implementing a high-performance NIC presents significant challenges. Both the development process and subsequent verification/performance optimization phases involve substantial technical complexities that demand considerable efforts. Specifically:

1) In the design and implementation phase, developers implement new protocols or algorithms as efficient processing pipelines. This requires not only ensuring the correctness of protocol functions during hardware coding, but also considering the efficient interaction between the NIC and other system components (i.e. PCIe bus, system memory). For instance, for memory access operations of some stateful protocols (such as RDMA, TCP, etc.), it is necessary to focus on the efficient interaction between the NIC memory hierarchy and the host memory subsystem (for example, reducing the additional PCIe traffic caused by internal cache misses of the NIC[16]);

2) During the verification and performance tuning phase, the interaction between the software driver, communication library, system bus and network card need to be considered comprehensively, which is far more difficult than the functional verification of an independent hardware unit. For example, when testing the performance of a RNIC, we need to carefully set the number of memory regions (MR), connections, Max Read Request Size (MRRS) of PCIe and other critical parameters [20][21].

Therefore, these two stages often require a lot of time and effort to verify the effectiveness of the new protocol. The end-to-end cycle—from initial protocol design to real-system validation—often spans multiple months or even years. As discussed in Section 1, the rapid evolution of application communication paradigms continues to drive innovation in network transport protocols. This creates pressing demand for a development framework that can simultaneously accelerate the development, implementation and verification of new protocols. Such a framework would empower researchers to rapidly prototype novel solutions and maintain architectural flexibility.

### 2.3 The Need for Rapid NIC Prototyping

We discuss in detail the requirements that we expect a NIC prototyping framework ought to meet and then discuss why existing works are hard to meet them.

#### 2.3.1 Requirements

Based on the imperative for rapid validation of novel network protocols, we have distilled the following core requirements for a NIC prototyping framework:

① Support for High-Performance Protocol Offloading

As previously discussed, general-purpose processors struggle to cope with the increasing network speeds, resulting in continuously shrinking instruction windows for packet processing. Achieving line-rate performance often requires dedicating dozens of CPU cores to network I/O operations, leading to significant computational resource and memory bandwidth waste. Consequently, modern innovative protocols increasingly adopt NIC hardware implementations for performance-critical operations. The NIC development framework must consequently provide architectural support for efficient implementation of novel protocol features at the microarchitecture level.

② Agile Development.

It is not that easy to shift those protocol processing task from CPU to NIC hardware, especially for those transport protocols which have complex memory access patterns or consume lots of memory. For connection-based protocols, we need to manage thousands of concurrent connections, which requires us to carefully allocate/deallocate the scarce on-chip SRAM. The process of designing, verifying and optimizing such a protocol-offloaded NIC usually takes great effort, which hinders the rapid iteration of new protocols and algorithms. We expect the framework to provide reusable and protocol-independent building blocks, allowing developers to implement new protocols with only minor modifications, rather than rebuilding a complete NIC from scratch.

③ System-Level Simulation.

Typically, a network simulator (OMNeT++[22], NS3[23], etc.) is used to verify the correctness and efficiency of a network protocol from a high-level perspective. This is fine, but not enough. If we want to dive deeply and make it clear how these protocols affect the interaction between the NIC and the system (for example, how on-NIC cache misses affect the message rate), we usually need to rely on tools (such as PCIe Logic Analyzer) to debug and analyze a real system, which is laborious and time-consuming. We expect that the NIC prototyping framework could accurately simulate the system characteristics, including the memory layout, and the behaviors of the driver and the system bus (e.g. PCIe).

### 2.3.2 Limitations of Current Solutions

We argue that current NIC prototyping frameworks cannot satisfy all the above requirements. They could be roughly divided into software-based and hardware-based solutions.

1) Software-based framework. These frameworks[24][25] are typically composed of a commercial ethernet NIC, a user-level driver and a communication library. The driver and library provide high-performance packet channels between applications and the underlying NIC, which avoids data copy between kernel space and user space. The burden of protocol processing is moved to application logic. It is easy to implement and deploy new protocols (goal ② and goal ③), but as we have stated before, most of the work is done by CPU, which violates the goal of line-rate processing (goal ①).

2) Hardware-based frameworks. Depending on the underlying hardware platform, these frameworks can be further divided into Smart-NIC[26][27] or FPGA-NIC[28][29]. The Smart-NICs are typically composed of clusters of ARM processors, which improve the programmability and flexibility of the framework, and meet the demand of agile development(goal ②), but they can hardly satisfy the performance requirements (goal ①). Typically, they can achieve about 10μs in terms of latency, tens of gigabits in terms of bandwidth, which is weaker than the performance of a protocol-hardened NIC(1~2μs, 100Gbps). FPGA-based frameworks can achieve nearly line-rate packet processing, but on the one hand, there are problems of difficulty in developing and debugging(goal ①). More importantly, the existing open-sourced FPGA NIC frameworks lacks support for stateful protocol offloading, specifically, no high performance on-chip buffer and system memory interaction mechanism is provided (see section 2.1). At the same time, most of these frameworks do not provide a system-level simulation platform (goal ③).

A comprehensive comparison of existing NIC prototyping framework is illustrated in **Table 1**, and no solutions could meet all three goals simultaneously (The most related work is Corundum[36] and RecoNIC[38], detailed discussion could be found in Appendix E). To fill this gap, we propose JingZhao, an open-source framework for rapid NIC prototyping. In the following sections, we will give a detailed description of its components, design choice and evaluation results.

**Table 1: Comparison of different NIC prototyping framework**

| Existing Solutions | Brief Description | Open-Source | Goals for rapid NIC prototyping framework | | |
|---|---|---|---|---|---|
| | | | ①: High Performance Offloading | ②: Support Agile Development | ③: System-Level Simulation |
| AlveoLink[30] | Provide IPs to build scale-out solutions on multiple FPGA evaluation boards. | 😐 | 😐 | 😐 | 😫 |
| XUP VNx[31] | Contains IPs that can be used to add 100Gbps networking to the design. UDP is used as the transport protocol. | 😐 | 😫 | 😫 | 😫 |
| Coyote[32] | A framework that offers operating system abstractions and a variety of shared networking, memory and accelerator services for modern heterogeneous platforms with FPGAs. | 😐 | 😫 | 😐 | 😫 |
| FpgaNIC[33] | An FPGA-based, GPU-centric NIC that provides reliable 100Gb network access to remote GPUs. | 😫 | 😐 | 😫 | 😫 |
| COPA[34] | Provides necessary networking/accelerator | 😫 | 😐 | 😐 | 😫 |
| | infrastructure and abstracts the underlying FPGA as a middleware. | | | | |
| QEP[35] | Provides queue-based DMA interface to facilitate NIC pipeline design. | 😨 | 😬 | 😨 | 😬 |
| Corundum[36] | A full-fledged ethernet NIC, accompanied with kernel driver; compatible to BSD socket interface. | 😬 | 😨 | 😨 | 😬 |
| Limago[37] | A full-fledged ethernet NIC supporting TCP/IP protocol offloading. | 😬 | 😬 | 😨 | 😨 |
| RecoNIC[38] | A framework provides IP blocks to build specific network functions. | 😨 | 😬 | 😬 | 😨 |
| JingZhao | A framework for rapid NIC prototyping that enables efficient hardware offloading and fast validation of network protocols. | 😬 | 😬 | 😬 | 😬 |

## 3 Overview of JingZhao

In this Section, we first give a brief overview of the architecture of JingZhao, and then discuss how can we meet the requirements of rapid NIC prototyping.

### 3.1 Architecture

As shown in Fig. 3, JingZhao is comprised of two main components: SHOST (Simulated Host) and PONIC (Protocol-Offloading NIC). They work collaboratively to simulate the interaction between the CPU, memory, system bus and the NIC. The SHOST part (implemented in System Verilog) contains several hosts, each host simulates the behavior of a NIC driver and communication library in a real system, including memory allocation and deallocation, on-NIC registers modification, etc. The PONIC part is comprised of several full-fledged NICs (implemented in synthesizable Verilog HDL), which are functional in a real system (see section 6.2 for our FPGA-based evaluation and Section 6.3 for our silicon validation). Each NIC is bound to a unique host (currently, we have not supported one-host-multi-NICs mapping, but it's easy to extend). The register space of the NIC is mapped into the host physical address space, hence the NIC could interact with the host through Programmed I/O and DMA, same as described in Section 2. Researchers can offload transport protocols based on the hardware templates (several subsystems and building blocks) provided by PONIC, then generate simulation stimuli within SHOST. SHOST will interact with PONIC through the simulated system bus to conduct precise evaluations of the network card's performance and functionality. Different NICs can be connected directly or through a simulated switch (not shown in the figure).

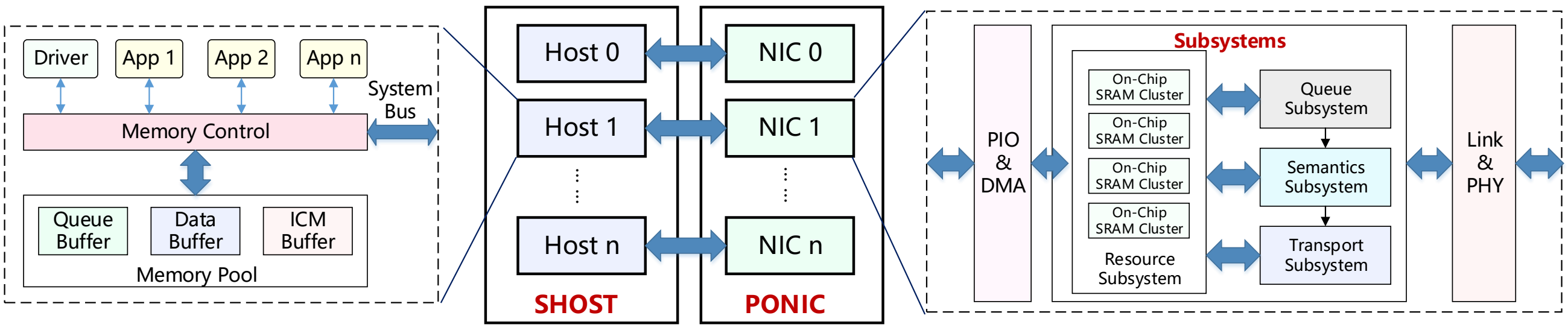

**Figure 3: Architecture of JingZhao Framework**

### 3.2 Design Philosophy

#### 3.2.1 Support for Agile Development

The core insight of our design stems from two critical observations:

1) From a functional perspective, a NIC works in a **4-W** manner: **Which** connection, at **When**, in **What** format, send **What** data. Implementing a new protocol typically requires modifying only subsets of these functional components while leaving the remainder of the architecture intact. For instance, a novel congestion control algorithm solely governs **When** to inject traffic into the network without altering packet formats.

2) The abstractions of current reusable IP blocks are too general, which cannot perfectly match the requirements of packet processing. For example, FIFO provides a single-queue abstraction to push and pop data, which is convenient, but not enough. For a connection-based protocol, we need to manage the states of connections individually, each connection should be mapped to a logical FIFO separately. Since all connections cannot be active simultaneously, a better way is to provide multi-queue semantics and manage these queues dynamically, which allocates and deallocates the queue buffer on demand, just as the malloc() / dealloc() function do in software (see section 4.1.3 Semantics Subsystem).

Based on these observations, our core idea is to provide several loosely coupled subsystems and protocol-independent building blocks, the functions of which are orthogonal and can be customized individually. Researchers could focus on developing and optimizing the part they are mostly interested in, and conveniently integrate the newly developed functions into the existing framework, reducing cost of time and labor work. The 4-W functional model are implemented as 4 subsystems, including Resource Subsystem, Queue Management, Semantics Subsystem, and Transport Subsystem, which could be leveraged to easily compose a functional, full-fledged high-performance NIC. The details of these subsystems are discussed in Section 4.

### 3.2.2 High-Performance Offloading

A high-performance NIC usually takes the following abstracted dataflow model. As shown in **Figure 4** (RX Pipeline is omitted), a PPU (Protocol Processing Unit, implemented as in hardware circuit) is responsible for a specific task, such as hash calculation, packet header modification, DMA read/write, etc. Each PPU accepts the inputs from its upstream PPUs, do some protocol-specific calculation or read/write memory, and then passes data to its downstream PPU, until the packet is injected into the network or written into memory. According to the data dependency defined by a protocol, several heterogeneous PPUs form a stage, and several stages construct the processing pipeline.

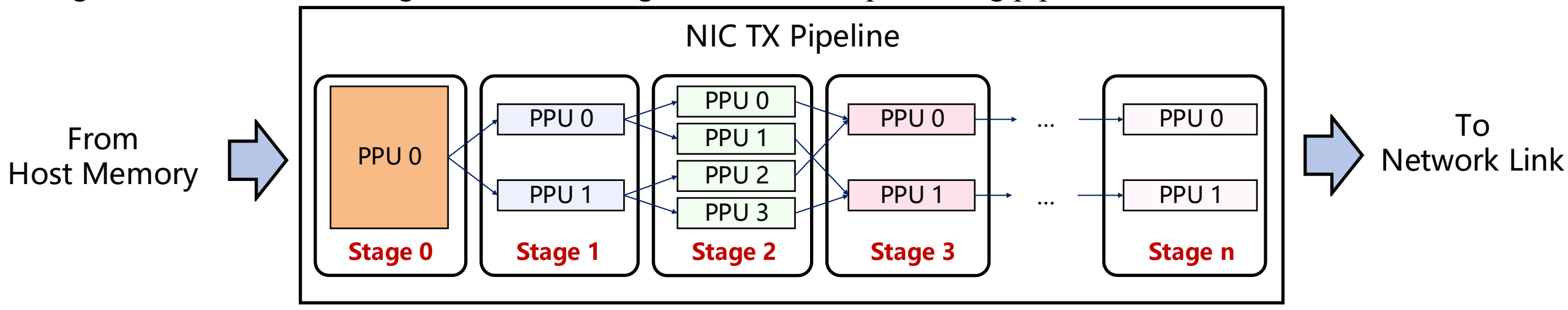

**Figure 4: Dataflow Model of a Processing Pipeline**

In this model, the sum of the maximum latencies of PPUs in each stage constitutes the total pipeline latency, while the minimum throughput of PPUs in any stage determines the overall pipeline throughput. For a stateless protocol (UDP, IP, etc.), it's easy to chain several building blocks together to make a pipeline functional, and improve the overall performance by replicating the PPUs in a single stage or directly replicating the pipeline. However, for a stateful protocol (RDMA, TCP, etc.), things become complicated, since several pipeline stages need to frequently read or modify some data structures. Even if there's cache on NIC, a cache miss still brings significant performance penalty, which is widely observed in commercial RDMA NICs. The reason behind this phenomenon is that a cache miss will trigger a system bus traverse and the pipeline is stalled until the desired data is fetched from host memory, we have also confirmed this with a NIC vendor. In this paper, we propose a non-blocking memory access mechanism which could amortize the cache miss penalty and reduce the negative impact on bandwidth (See Section 6.1.2.2), and it could be generally applied to stateful protocol offloading.

### 3.2.3 System-Level Simulation

The goal of system-level verification is accurately simulating the interaction between a NIC and a real system. It's impractical to fully simulate a real system, so we have to trade-off between accuracy and complexity.

1) For CPU simulation, we just preserve its PIO behavior, which modifies the on-NIC registers through simulated system bus. Applications running on CPU are simplified as several network flows, which send/receive data to/from network in a predefined manner;

2) For memory simulation, we do not implement any complex mechanism of memory management, since in the view of NIC, it does not care about how physical memory is organized. However, in order to support some protocols allowing virtual-address-based memory access, we implement a naïve virtual-to-physical address mapping strategy (See Appendix B);

3) For system bus simulation, since the NIC directly talk to system bus and this requires 1:1 simulation, we do implement the transaction layer of a typical system bus (PCIe).

The details of each component will be discussed in the following Section.

## 4 Design

The hardware architecture of JingZhao NIC is derived from the 4-W functional model, and is composed of 4 loosely coupled subsystems, including Resource Subsystem (RS), Queue Subsystem (QS), Semantics Subsystem (SS), and Transport Subsystem (TS), each subsystem is composed of several pipeline stages, following the functional model described in Section 3.

It is worth noting that Semantics Subsystem decides in "What" format the network data should be transferred. The key difference between this subsystem and others is that SS will not only modify packet header, but also touch packet payload for protocol-dependent or application-specific packet processing. For example, a RDMA NIC and an Ethernet NIC could share the control logic of the QS (with just a few modifications to the hardware data structures, e.g. Queue Status, etc.) for queue scheduling and the TS for reliability control, but they differ a lot in packet format, memory access pattern and the semantics offered to applications. Therefore, our key idea is to standardize the framework of QS, RS and TS, which helps researchers focus on the design of protocol- or application-specific packet processing logic. We will describe these subsystems in detail and show concrete examples of how to implement a new protocol by using these building blocks.

To use JingZhao framework to design and verify a domain-specific NIC, the users need to do three things: 1. Specify the parameters of common building blocks (Queue/Resource/Transport Subsystem); 2. Developing protocol-dependent hardware logic using HDL (Semantics Subsystem provides several commonly used primitives which could be instantiated as hardware modules); 3. Chain these blocks together to build a pipeline and connected to the System Bus in SHOST (if FPGA prototype is required, the pipeline will talk to CPU and system memory via a PCIe controller provided by FPGA commercial tools, i.e. Vivado).

## 4.1 Design of PONIC

### 4.1.1 Resource Subsystem (RS)

Resource Subsystem manages communication resources which are frequently accessed by different pipeline stages. It provides read/write interface (ResChnl) to operate on protocol-dependent data structures. These data structures are created and allocated in host memory, and managed by RS. RS maintains a Resource Buffer to cache these data structures. Resource read request is handled by CacheRead while resource write or delete is handled by CacheModify. Requests/Responses from/to different pipeline stages are muxed/demuxed. Depending on the characteristics of specific protocols, the logic of RS could be replicated to manage different resources. The overall architecture of RS is shown in **Figure** 5(a).

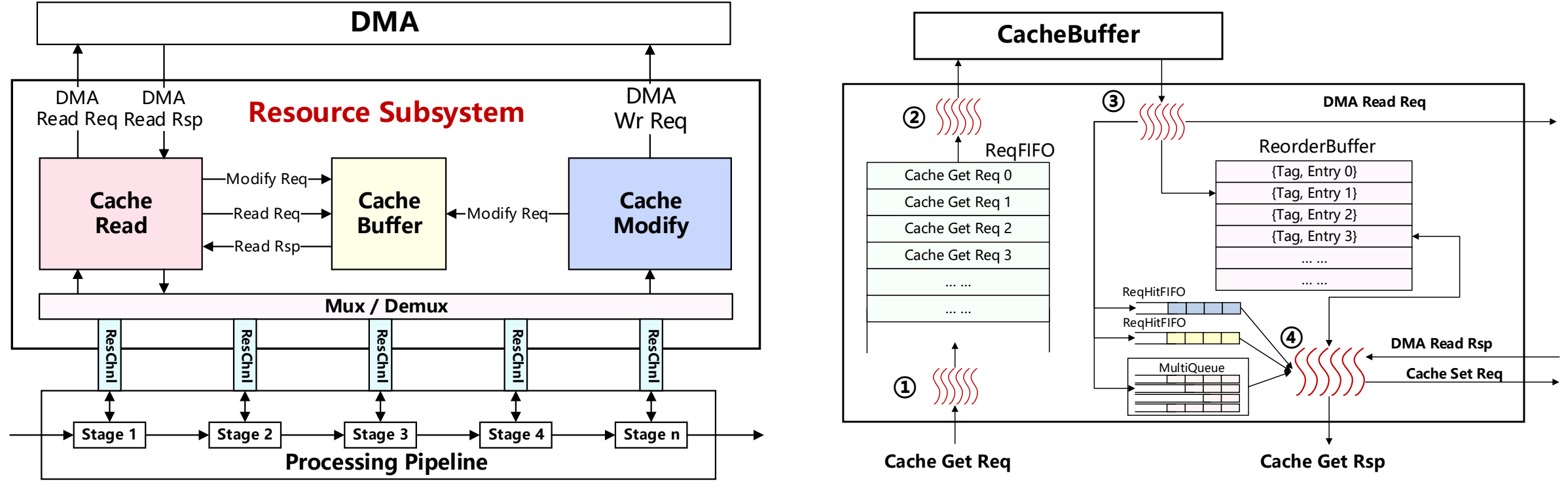

**(a) Interaction between RS and other Subsystems** **(b) Non-blocking Resource Access**

**Figure 5: Architecture of Resource Subsystem**

For most connection-based stateful protocols, the cache requests/responses belong to the same connection should be issued/returned in the same order, while requests/responses belong to different connections could be interleaved. If a request encounters a cache miss, it will issue a DMA Read to fetch desired data from host memory and then update cache. Without proper design, all the following requests may be blocked by the DMA request no matter whether they belong to different connections or whether their desired data resides in cache. The phenomenon is similar to the HOL (Head-of-Line) problem in switch design, and discussed earlier in[18], as shown in **Figure** 6.

To solve this problem, we have adopted a VoQ-like design, as shown in **Figure** 5(b). The Multi-Queue block (see Section 4.1.3) is the key component, which provides Enque/Dequeue interface for different connections. Four hardware units work collaboratively to deal with cache hit and cache miss. Based on the result of cache lookup, unit ③ will push a signal into ReqHitFIFO or ReqMissFIFO, and insert the request into a logical queue of the Multi-Queue block. If there is a cache hit, the cached data is stored into a ReorderBuffer for further processing, otherwise, a DMA Read is issued to system memory. Unit ④ is triggered by the signal of ReqHitFIFO, ReqMissFIFO or DMA Read Response. Since the requests of different connections are pushed into different logical queues, Unit ④ can deal with their responses separately. A cache miss will only block the requests of the same connection, and will not affect other connections. The size of reorder buffer should be larger than the bandwidth-delay-product of the system bus to ensure that even if all the requests encounter cache miss, the bandwidth of the pipeline will be slightly affected.

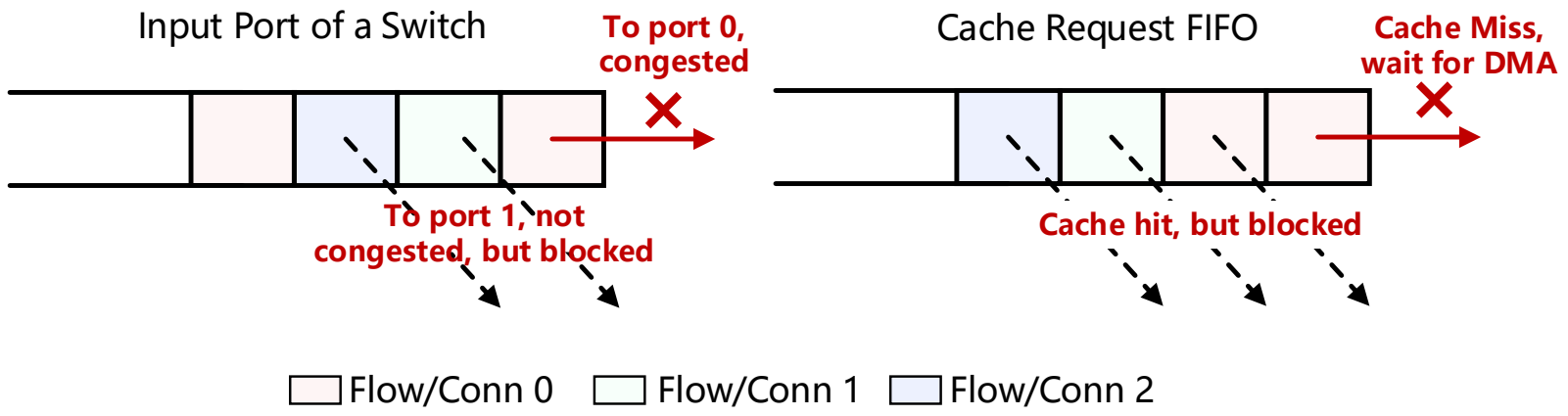

**Figure 6: Problem of Head-of-Line Blocking**

### 4.1.2 Queue Subsystem (QS)

The NIC is abstracted as multiple SQs, RQs and CQs, as described in Section 2. As shown in **Figure** 7, each queue is accompanied with several data structures, which describe the working status of the queue. A Queue Cache is adopted to avoid frequently fetching WQEs from host memory. The Cache Buffer is divided into several Cells, each Cell is composed of several Slots, whose width are aligned to the segments of WQEs (segment is the minimum unit to be processed by WQE Fetcher and WQE Parser). A cell is owned by a specific Queue, and could be occupied by another Queue if needed. We use lower bits of Queue Number to map different Queues to these cells. A Cache Record is used to indicate the offset (points to start of a valid WQE in the cell) and the owner of the Cell, which helps WQE Fetcher determine where to address a WQE. Each time a WQE is fetched, the offset will be increased. If the number of available WQEs is lower than a predefined threshold, i.e., the offset of a cell is reaching the end of the cell, Queue Cache will automatically DMA Reads WQEs from host memory, refill the Cache Buffer and update the Cache Record. A Rate Limiter is adopted to facilitate some congestion control algorithms. It tracks two data structures of each queue, which collaboratively indicate how much data can a queue continue to send. In current design, we do not implement any CC algorithm, but provides an interface to modify these data structures. It is worth noting that, all the above data structures could be integrated into the Resource Subsystem to leverage the optimized Cache-Host Memory hierarchy for better scalability, except for the Queue Cache, since the prefetching behavior of Queue Cache is tightly coupled with the queue abstraction, we will leave this integration for future work.

The Queue Subsystem works as follows. When software rings the doorbell through PIO, the Queue Scheduler looks up and modifies the status of the queue designated by the doorbell, and then notifies the next pipeline stage. The WQE Fetcher will get a WQE from Queue Cache. The structure and size of a WQE is protocol-specific and could be configured. The WQE is then passed to the WQE Parser to be decoded and translated into several metadata, indicating the source/destination address of a packet, operation code, memory address of network data, etc.

These metadata are forwarded to Semantics Subsystem for further processing. During the process of parsing a WQE, the Rate Limiter will tell WQE Parser how much data could be injected into the network. If there still exists available WQEs in the Send Queue (i.e. Head is not equal to Tail), the Queue Number will be passed to Queue Scheduler for next-round scheduling. The above process will be repeated until all the Send Queues are cleared.

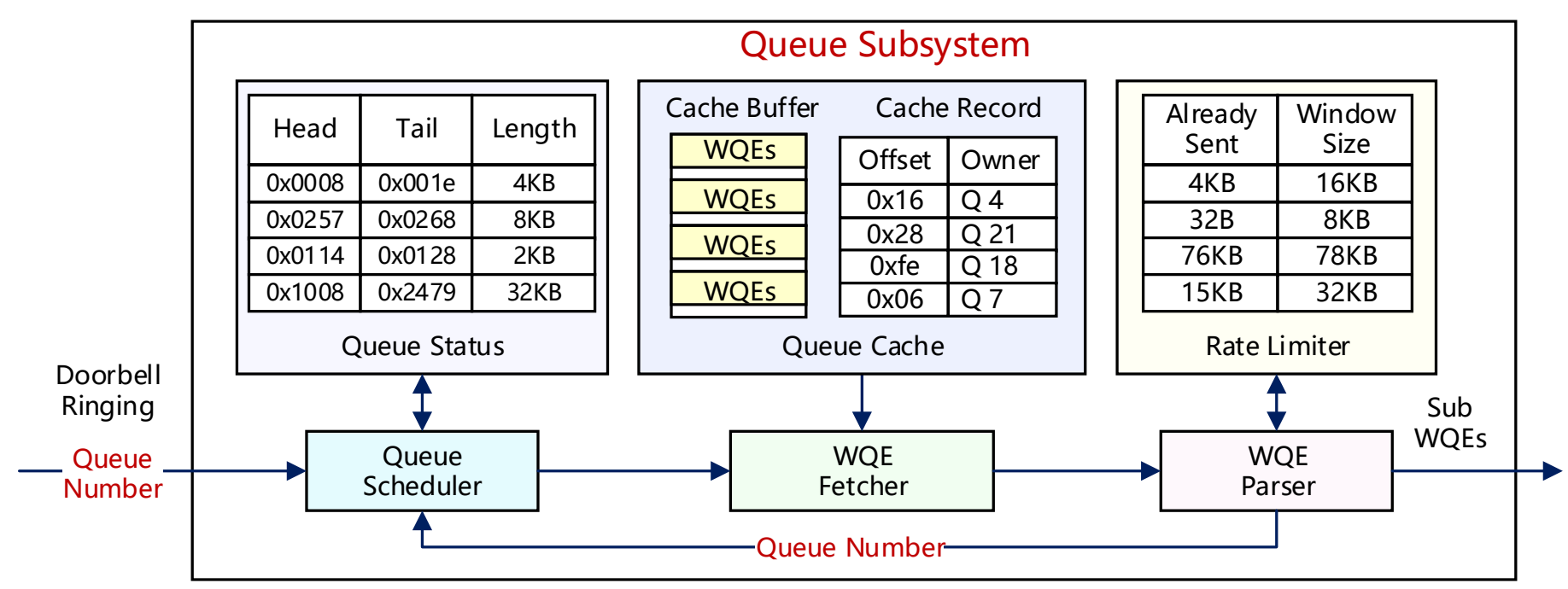

**Figure 7: Architecture of Queue Subsystem**

### 4.1.3 Semantics Subsystem (SS)

Semantics Subsystem are responsible for application- or protocol-specific functions, hence we leave the SS development for designers. Several frequently used network primitives are provided to facilitate designing and RTL coding. **Table** 2 illustrates the functions, interfaces, and use cases of different primitives. All these primitives are implemented as independent hardware IP blocks which could be configured and instantiated on demand. At the same time, we are continuing adding new primitives into this library, we hope to make this library the "Glibc" of NIC development. Detailed discussion and usage of these primitives could be found in Appendix A.

**Table 2: Abstracted Hardware Primitives for NIC Development**

| Hardware Primitive | Function | Use Case |
| --- | --- | --- |
| Append Header | Append header fields to a packet. | Encapsulate payload into a packet. |
| Remove Header | Remove header from a packet. | Decapsulate payload from a packet. |
| Scatter Data | Scatter a packet to different buffers of host memory. | Support for non-contiguous memory access, reducing the overhead of copying data into contiguous network buffer. |
| Gather Data | Gather data from non-contiguous buffers in host memory and encapsulate them into a packet. | |
| Dynamic Enqueue/Dequeue | Push/pop data into/from a logical queue in a shared buffer. | Map different connections to logical FIFOs, which could share a block RAM. |
| Dynamic Insert/Delete | Allocate and deallocate space in a shared buffer. | Dynamically manage buffer space, like malloc/dealloc function. |

### 4.1.4 Transport Subsystem (TS)

Transport Subsystem provides end-to-end reliability to the Semantics Subsystem. For a reliable protocol, the SS just injects packets into TS, and has no burden to care about loss detection and retransmission. If an unrecoverable error happens (for example, link failure), an event is generated by TS to notify SS deal with the error. For an unreliable protocol, the SS could directly talk to network link, bypassing the TS.

- **Go-Back-N (GBN) or Selective Repeat (SR)?**

It is difficult to definitively state which reliability algorithm is superior. Selective Retransmission (SR) achieves higher packet loss recovery efficiency at the cost of increased complexity and additional memory overhead (for reordering out-of-order packets), while Go-Back-N (GBN) consumes fewer SRAM resources but retransmits significantly more redundant data. We provide both algorithms as fundamental building blocks. Designers can select different algorithms for integration into the NIC framework based on specific use cases. SR is set as the default algorithm because in most scenarios—particularly when network switches' adaptive routing functionality is enabled, generating substantial out-of-order packets—we prioritize performance over resource consumption. However, if we have little budget for chip area, GBN is preferred (See **Table** 7 for hardware resource consumption). The architecture of TS is shown in **Figure** 8.

The GBN and SR algorithms share several building blocks, such as InOrderInject on TX path, which appends PSN to a packet, inserts the packet into a packet buffer for further retransmission, and injects the packet into network; and InOrderCommit on RX path, which commits packet to the SS in the same order as the packets are sent. The algorithm of retransmission and out-of-order detection has been widely studied by previous works[12][13][14], and we will not discuss in detail.

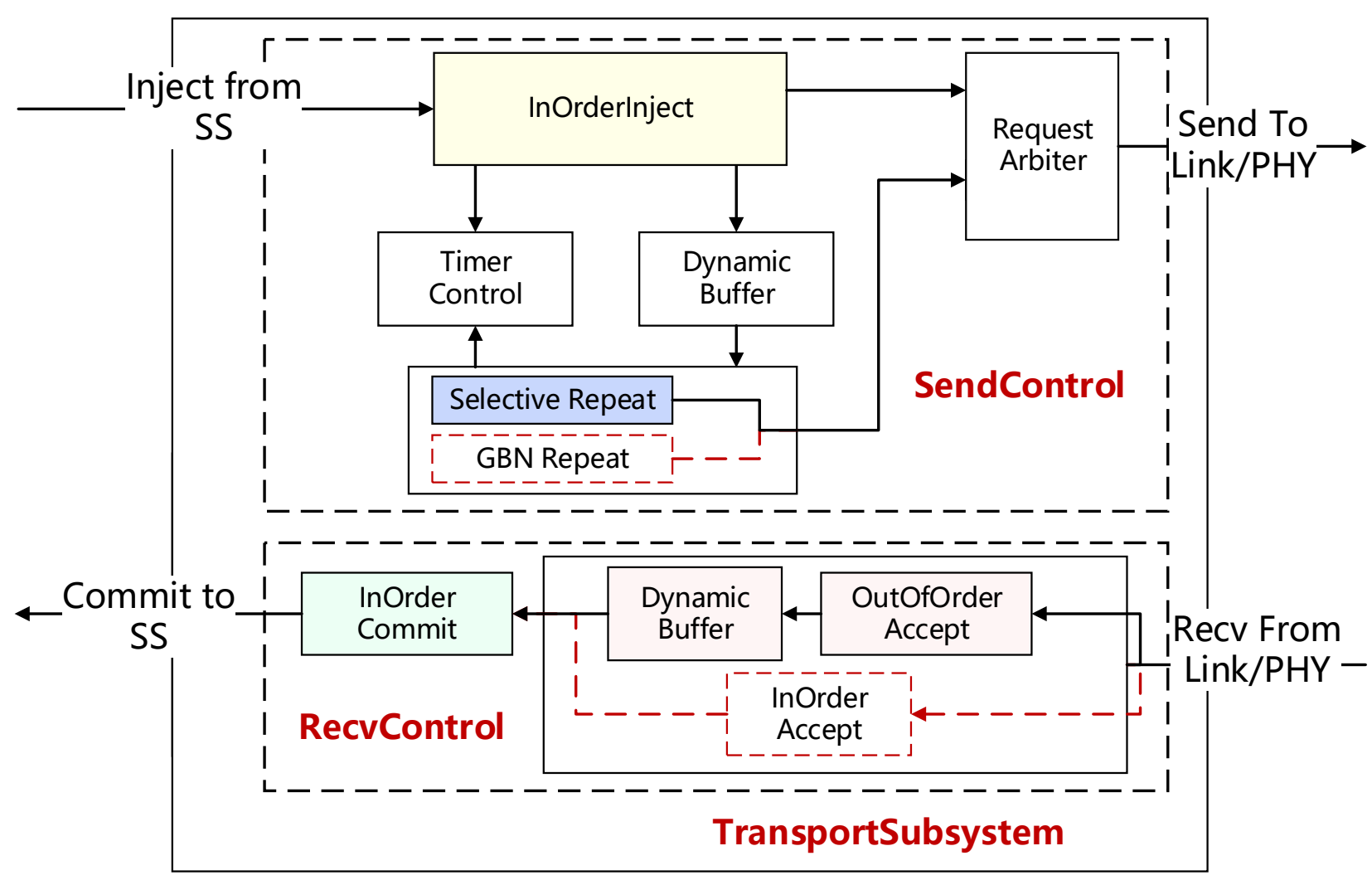

**Figure 8: Architecture of Transport Subsystem**

### 4.2 Design of SHOST

As shown in **Figure 3**, each host consists of five main components: the Application, Driver, Memory Control, Memory Pool, and System Bus. The Application layer can enqueue customized WQEs (see Section 2.1) into Send Queues and update the doorbell registers on the NIC. In the current design, we have not implemented any complex computational logic, the Application is treated as a network traffic generator that produces WQEs in a predefined manner. The Driver partitions the Memory Pool into different regions, including Queue Buffers, Data Buffers, and ICM Buffers, mirroring the structure found in real-world systems. It provides the Application with a queue-based abstract interface and encapsulates several NIC control interfaces. The Memory Control offers read/write interfaces for both the Driver and the NIC, intercepting DMA requests issued by the NIC hardware and translating them into accesses to the Memory Pool. The System Bus simulates data transfer between the NIC and the system. The current version implements a PCIe bus component that performs some functions of the PCIe transaction layer.

### 4.3 Put Them Together: Example Design

In this Section, we will give examples of two specific designs to show how to integrate protocol-specific functions into our framework.

#### 4.5.1 A Full-Fledged RDMA NIC

RDMA NIC is widely used in High Performance Computing systems in the past few decades, and is now becoming favored by Datacenter operators since it provides ultra-low latency and high bandwidth, which is desirable for large-scale distributed applications, such as Big Data and LLM training. To implement a RDMA NIC, we make several modifications to Queue Subsystem and Resource Subsystem, and develop a RDMA Core based on the primitives provided by the Semantics Subsystem to interact with standard OFED libibverbs.

1) Minor modifications to Queue Subsystem. RDMA protocol allows a WQE contains multiple memory address, hence the length of WQE is variable. To support this requirement, we modify the WQE Fetcher to read fixed-length data block from Queue Cache, which ensures that an entire WQE is passed to next stage. The internal state machine of WQE Parser is modified to support multiple memory address decoding, and it will pass these addresses as metadata to the RDMA Core in a stream manner.

2) Configure the Resource Subsystem. RDMA processing needs to access four types of communication resources, including QP Context, CQ Context, Memory Protection Table and Memory Translation Table. These resources are accessed individually, hence each type of resource is instantiated as a Resource Subsystem, providing CacheRead/CacheWrite interfaces to RDMA Core. The data widths of these interfaces are aligned to the requirements of specific resources (416-bit for QPC, 128-bit for CQC, 256-bit for MPT and 64-bit for MTT).

3) Developing a RDMA Core. It is composed of 4 sub-cores, including ReqTransCore, ReqRecvCore, RespTransCore and RespRecvCore, responsible for transmitting/receiving request/responses. Each core is composed of several pipeline stages. The functions of stages could be abstracted as the following repeated operation sequence: <read resource -> generate a header field-> gather/scatter data -> append a header field>. As we have described in Section 4, since most of these actions are generalized as building blocks, the work of development is mainly generating the header field and DMA read/write data from/to host memory. **Figure 9** shows the pipeline stages of ReqTransCore, and we see that most of them could be directly mapped to the existing primitives. Only several hundreds of RTL codes need to be done.

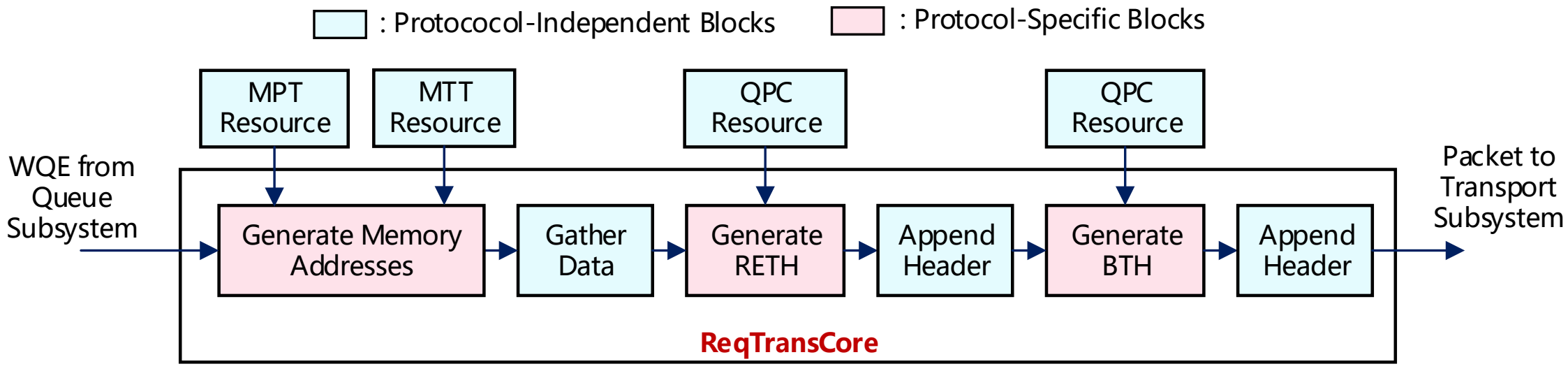

**Figure 9: Architecture of ReqTransCore of the RDMA Core**

#### 4.5.2 An In-Network Key-Value Cache

Key-Value Store is a commonly used method to store unstructured data. Typically, a client sends a key to the server, the server lookups its local database according to the key, and then returns the value. The throughput of the system is bounded by the hash calculation and memory traverse. Several systems have been proposed to offload Key-Value Store to NIC hardware to accelerate the process[41][42]. We design a simple Key-Value core, which can be easily integrated into the NIC framework.

Different from RDMA core, KV Core is much more application-specific, it contains many actions that cannot be paralleled, for example, a SHA-256 hash calculation of the key needs 64 cycles to output a result. According to the pipeline model in Section 3, we instantiate several Hash Calculation cores to balance the throughput of different pipeline stages. The architecture of the KV Store is shown in **Figure 10**. The payload is appended with an Ethernet header. The WQE contains a Key which is generated by software. The Resource Subsystem is instantiated as Value Memory to store the key-value pair. Since the Cache Buffer in Resource Subsystem is divided into fixed-length slot, in current design, each key-vale pair is fixed-length and fit into the Cache Buffer.

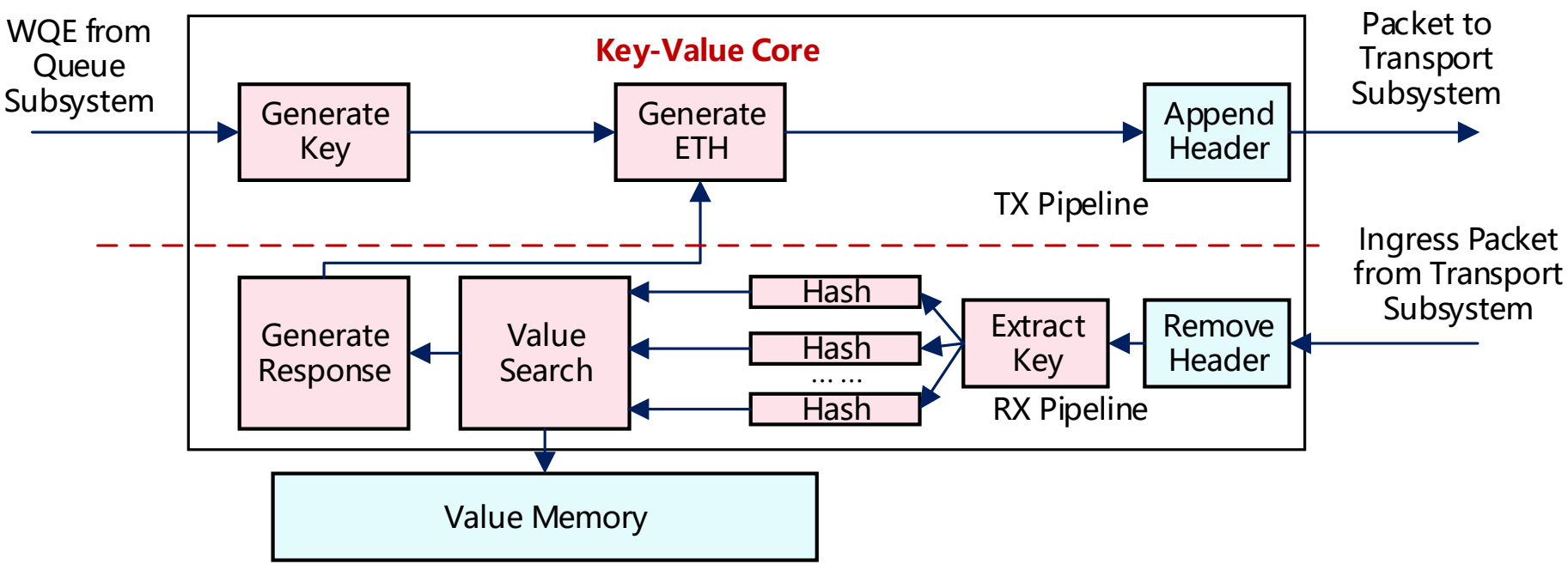

**Figure 10: Architecture of a In-Network Key-Value Cache**

## 5 Implementation

This Section describes the implementation details of our framework.

### 5.1 SHOST

The SHOST part of JingZhao is implemented using System Verilog in about 15000+ LoCs, and integrated into the UVM verification framework. The Application, Driver, Memory Control, Memory Pool and System Bus are implemented as different UVM classes. The System Bus provides four AXI-Stream interfaces to talk to the PONIC (CQ interface to write/read on-NIC registers/buffer, CC interface to accept response from the NIC; RQ interface to accept memory requests from the NIC and RC interface to return responses to the NIC).

### 5.2 PONIC

All of the NIC hardware building blocks are developed using Verilog HDL in about 20000+ LoCs. To make PONIC interact with SHOST and real system, we have also implemented a DMA Engine, which provides simple memory read/write semantics to other blocks (for instance, Resource Subsystem issues a <read, address, length> command to DMA Engine, and DMA Engine will automatically generate PCIe read requests). Details of the PCIe transaction are hidden by the DMA Engine, including payload splitting, out-of-order Cpl, etc. To measure the resources consumption of our implementation, we have also built a full-fledged RDMA NIC using a Xilinx VCU128 FPGA evaluation board, which is embedded with an UltraScale+ VU37P chip. To interact with the host and network, we instantiate a UltraScale PCI Express 4c Integrated Block. The reference clock of the NIC is 50MHz. The resource usage is shown in Appendix F.

## 6 Evaluation

This section evaluates the JingZhao framework through three key aspects: 1) System-level simulation. The SHOST emulates real-system stimuli to comprehensively assess the functionality and performance of each NIC component; 2) FPGA prototype validation. We run Perftest[43] benchmark to test our implemented open-source RDMA NIC in real systems. Based on JingZhao framework, we have successfully taped out a chip using TSMC's 28nm process, the results of silicon validation could be found in Appendix D.

### 6.1 Simulation Results

#### 6.1.1 Microbenchmarks for Building Blocks

As described in Section 3, the performance of a pipeline is determined by the performance of each pipeline stage, hence, we first measure the bandwidth, throughput and latency of different building blocks. For each block, we generate continuous requests and payload to the input ports, and vary the size of the payload to measure the performance under different workloads. The frequency of the reference clock is set to 200Mhz. For each block, the internal data bus is set to 512-bit. Other block-specific configurations are as follows. For DMA Write/Read, we set Max Read Request Size and Max Payload Size to 512B, and the measured latency does not include that of the system bus; for Append/Remove Header, the width of the header is 8B; for Scatter/Gather Data, each request is composed of two packets. **Table** 3 summarizes the performance results.

**Table 3: Performance of Different Building Blocks**

| | Bandwidth (Gbps) | | | | | Throughput (Mops/s) | | | | | Latency (ns) | | | | |
|---|---|---|---|---|---|---|---|---|---|---|---|---|---|---|---|
| | 64B | 128B | 256B | 512B | 1024B | 64B | 128B | 256B | 512B | 1024B | 64B | 128B | 256B | 512B | 1024B |
| **DMA Write** | 52.50 | 70.35 | 84.96 | 93.45 | 96.60 | 62.50 | 38.46 | 27.78 | 17.86 | 10.42 | 166.12 | 176.49 | 186.50 | 206.99 | 246.84 |

| | | | | | | | | | | | | | | |
|---|---|---|---|---|---|---|---|---|---|---|---|---|---|---|
| **DMA Read** | 45.12 | 65.00 | 80.27 | 89.17 | 95.36 | 61.30 | 38.72 | 26.68 | 17.55 | 10.10 | 147.25 | 184.39 | 180.47 | 204.62 | 255.08 |
| **Append Header** | 85.98 | 87.92 | 92.36 | 92.49 | 94.89 | 90.93 | 47.58 | 31.27 | 19.38 | 10.99 | 11.35 | 21.02 | 31.88 | 51.55 | 91.20 |
| **Remove Header** | 88.23 | 86.59 | 93.48 | 93.80 | 95.64 | 91.18 | 45.62 | 33.30 | 19.47 | 11.78 | 11.37 | 21.04 | 31.71 | 51.83 | 91.36 |
| **Scatter Data** | 51.39 | 66.32 | 86.87 | 88.89 | 96.12 | 93.47 | 47.32 | 34.97 | 17.65 | 12.83 | 11.28 | 21.39 | 31.63 | 51.76 | 91.87 |
| **Gather Data** | 50.27 | 66.58 | 87.94 | 87.02 | 95.58 | 95.28 | 47.78 | 33.28 | 18.32 | 11.05 | 10.59 | 24.50 | 30.84 | 50.29 | 90.58 |
| **Cache Read** | 51.14 | 56.77 | 80.12 | 85.69 | 92.38 | 39.12 | 28.17 | 21.98 | 15.27 | 9.88 | 25.48 | 35.78 | 45.51 | 65.47 | 105.51 |
| **Cache Write** | 48.28 | 53.20 | 82.45 | 85.38 | 93.46 | 37.24 | 29.83 | 22.37 | 15.36 | 10.92 | 25.39 | 35.42 | 45.49 | 65.83 | 107.86 |
| **Dynamic Enqueue** | 39.83 | 43.41 | 53.38 | 66.67 | 96.49 | 39.83 | 29.75 | 20.82 | 15.49 | 9.41 | 25.13 | 35.83 | 45.70 | 65.98 | 105.50 |
| **Dynamic Dequeue** | 39.66 | 45.20 | 52.46 | 65.20 | 97.51 | 39.96 | 26.99 | 19.74 | 16.72 | 8.97 | 26.37 | 37.92 | 46.03 | 66.36 | 108.74 |
| **Dynamic Insert** | 39.49 | 41.38 | 57.14 | 72.83 | 95.28 | 48.78 | 28.36 | 24.69 | 16.53 | 9.95 | 20.59 | 30.50 | 40.66 | 60.09 | 100.50 |
| **Dynamic Delete** | 37.08 | 45.79 | 59.62 | 72.46 | 95.80 | 47.62 | 29.26 | 24.39 | 16.39 | 9.90 | 21.98 | 31.64 | 41.72 | 61.41 | 101.49 |

As for bandwidth, all the blocks can achieve nearly line-rate data processing at 1024B. For small and medium payload size, due to complex transition of state machine, some blocks can hardly operate at a high speed. For example, as for DMA Read, it is responsible for deal with the out-of-order Cpls. As for Dynamic Enqueue, it must allocate available space, update head/tail pointer before the payload is inserted into the buffer, occupying 3 clock cycles. There exist such serial actions which can not be overlapped with payload transfer, thus wasting available bandwidth. The similar performance characteristics can be observed for throughput. Since the width of the data bus is 512-bit, the throughput of a 64B packet is theoretically equal to the clock frequency, i.e. 200Mops/s. However, due to those serial actions, the maximum throughput of these blocks is 95.24Mpps (Gather Data), which includes an intermediate state to pop the metadata from a FIFO. To increase bandwidth and throughput of a pipeline state, we can simply instantiate multiple heterogeneous blocks in the same pipeline stage. Different connections or flows could be mapped to these blocks to achieve higher aggregated performance. As for latency, most of the blocks consumes less than 25ns to process a small packet, except for DMA Read/Write. The reason is that DMA directly interacts with system bus, hence it must implement the transaction layer protocol, which usually contains complex state machines. In our current implementation, a PCIe transaction typically needs ten or more state transitions, incurring extra latency, which will be optimized in the future.

The performance of two retransmission algorithms under different packet loss rate is also measured. At a relatively low loss rate ($<10^{-4}$), both Go-Back-N and Selective Repeat behave well as the bandwidth is about 95Gbps, close to the peak (set to 100Gbps). However, things become different when loss rate reaches $10^{-3}$, GBN drops sharply to 25Gbps. We have observed lots of retransmitted packets during the network, much more than that of SR. It is recommended to use SR logic for better performance, at the cost of a bit more memory consumption.

### 6.1.2 Evaluation for a Full-Fledged NIC

In this Section, we test the RDMA NIC and In-Network Key-Value Store using the simulation framework described in Section 3. In both tests, we connect two NICs using an AXI-Stream interface, one NIC is initiated as a Requester, and another NIC as a Responder. The Simulation part creates 256 Queues to interact with the NIC simultaneously. Each App is mapped to a specific Queue, and continuing pushing WQEs into the queues. For RDMA NIC, we test the basic performance (no cache miss) of RDMA Write and RDMA Read primitives, as well as the worst case (100% cache miss). The “no cache miss” case is formed by residing all the data structures in a larger cache buffer, while the “100% cache miss" scenario is formed by removing the cache, hence each resource access will traverse the system bus. The bandwidth, and throughput of the system bus are set to 100Gbps and 200Mops, which will not be the bottleneck of the communication. And the latency of the system bus is set to a PCIe RTT, which is about 350ns. For Key-Value Store, we test the throughput under different configurations of hash cores.

#### 6.1.2.1 Performance of RDMA NIC

**Figure 11**(a)(b)(c) shows the performance of RDMA Write/Read in different scenarios. For bandwidth, we see both Write and Read achieve high performance when there’s no cache miss. The peak bandwidth reaches 94.37Gbps for Write and 93.36Gbps for Read when the packet size is 4096B. Cache miss affect the bandwidth performance slightly. When cache miss happens, the Resource Subsystems DMA Reads resources from the ICM buffer in host memory, this behavior will compete for DMA bandwidth. The total size of communication resources is 108B (416-bit QPC, 128-bit-CQC, 256-bit MPT and 64-bit MTT), that’s to say, if there is no cache on NIC, each time we DMA Read 4096B payload from host memory, there will be an extra 108B overhead. The bandwidth loss is about (108/(108+4096))= 2.5%, which is acceptable. For throughput, things become different. When no cache miss happens, the throughput of small packets could reach a peak performance of 39.23 Mops/s, which is limited by the minimum throughput of the building blocks (**Table** 3). However, when cache miss happens, the throughput drops to 13.4 Mops/s, the reason is that when a NIC DMA reads a small packet from host memory, it also issues DMA reads for QPC, CQC, MPT, and MTT, and these requests equally share the throughput of DMA Read. Since the throughput of DMA Read is 45.12 Mops/s, the available throughput for packet is about 9.02 Mops/s. The similar issue could be observed in latency test. When there is no cache miss, the latency of a packet is about 2.9µs, while a cache miss incurs extra system bus traverses, and increases the latency by about 400ns. We leave this optimization to future work.

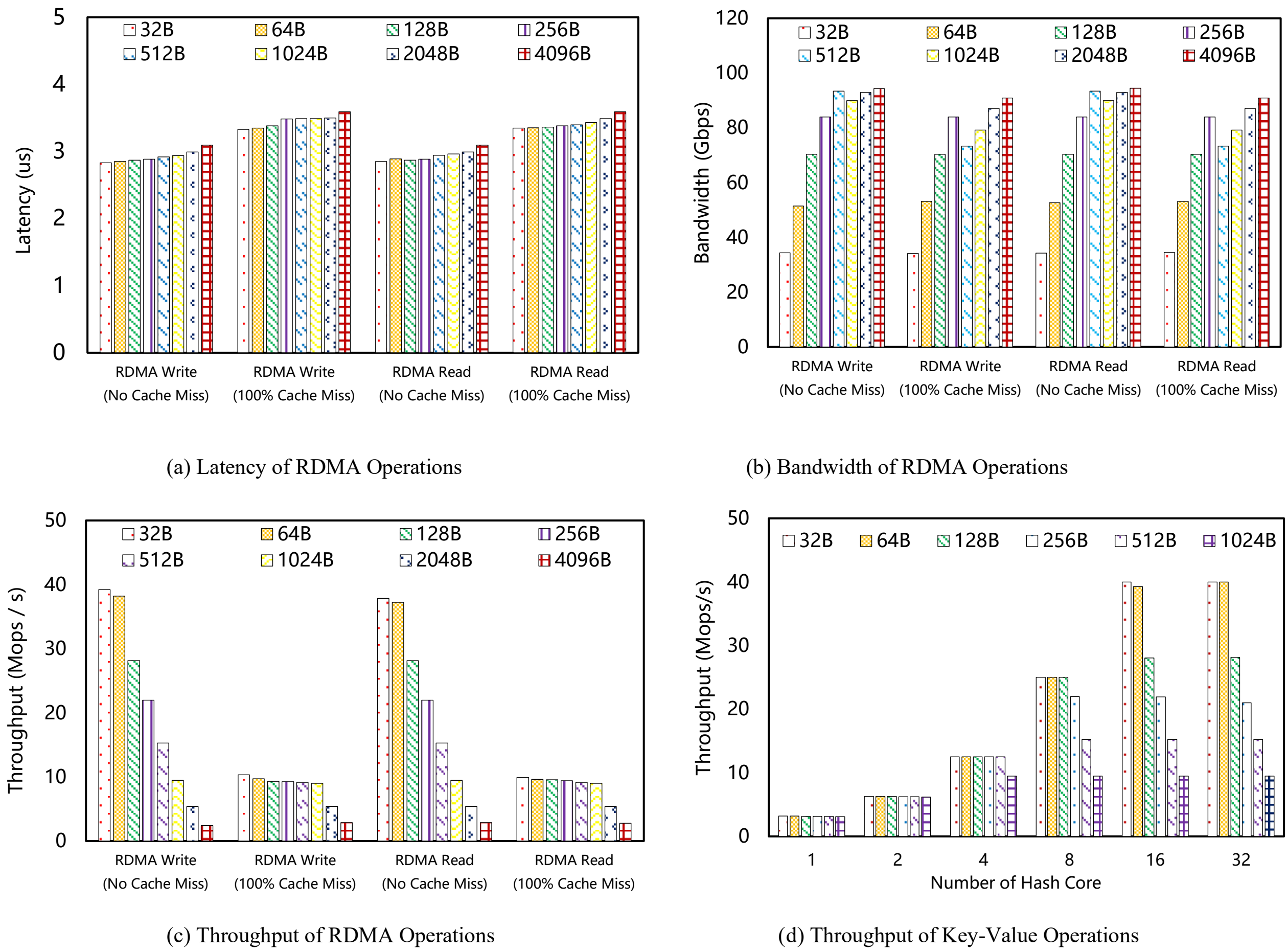

(a) Latency of RDMA Operations

(b) Bandwidth of RDMA Operations

(c) Throughput of RDMA Operations

(d) Throughput of Key-Value Operations

**Figure 11: Performance Evaluation for RDMA Core and Key-Value Cache**

#### 6.1.2.2 Performance of In-Network Key-Value Store

The maximum of throughput of the Key-Value Store pipeline is bounded by the minimum throughput of the hash cores and the other blocks (**Figure 11**(d)). A single hash core generates a 64-bit hash value every 64 cycles, providing a throughput of about 3.13 Mops/s, when there is only one hash core, it is the bottleneck of the system. As we have increased the number of paralleled hash cores, the throughput increases linearly, until reach the peak throughput of other blocks. 16 hash cores theoretically provide about 50Mops/s, while 32 hash cores provide about 100Mops/s, however, the overall throughput of the Key-Value processing pipeline is limited by the slowest Resource Management, which is 39.28 Mops/s.

### 6.2 FPGA Prototyping

In this Section, we test the functionality and performance of our RDMA NIC prototype. As discussed in Section 5, the a full-fledged RDMA NIC is implemented on Xilinx VCU128 evaluation board. A kernel-space driver and a user-space driver is implemented to make the NIC able to talk to the standard OFED libibverbs (detailed system architecture could be found in Appendix C). The tests are conducted using Perftest tools (by launching ib_write/read_bw and ib_read_lat). The FPGA platform runs at a clock frequency of 50Mhz, and the internal bus width is 512-bit, resulting in a theoretical bandwidth of 25.6Gbps. For each test, two nodes are connected. Each node is equipped with a Intel Core i9-9900K CPU, 32GB system memory, and 256GB SSD. The FPGA card is plugged into the PCIe slot of the motherboard, and connected to each other via an optical fiber. The performance results are shown in **Table 4**. The bandwidth performance is as expected, which is close to the theoretical results. The peak bandwidth can reach 23.5Gbps. The latency is higher than our simulation results. The extra latency (~0.5µs) mainly comes from the software running on the host, which is not measured in our simulation evaluation.

**Table 4: Performance of FPGA-Based RDMA Prototype**

| | RDMA Write | | RDMA Read | |
|---|---|---|---|---|
| | Bandwidth (Gbps) | Latency(μs) | Bandwidth (Gbps) | Latency(μs) |
| 32B | 4.92 | 3.16 | 3.16 | 3.25 |
| 64B | 6.87 | 3.19 | 5.62 | 3.29 |
| 128B | 11.33 | 3.24 | 10.08 | 3.46 |
| 256B | 19.75 | 3.27 | 18.50 | 3.82 |
| 512B | 22.11 | 3.31 | 20.86 | 3.99 |
| 1024B | 22.9 | 3.47 | 21.75 | 4.17 |
| 2048B | 23.25 | 3.80 | 22.33 | 4.40 |
| 4096B | 23.50 | 4.46 | 23.18 | 4.86 |

## 7 Related Work

### 7.1 Fast NIC Prototyping

Several works are proposed to provide NIC prototyping frameworks to facilitate network development. AlveoLink[30], XUP VNx[31], Coyote[32] are FPGA-based platforms, which offload the transport layer to NIC hardware. They offer low-level hardware interfaces to inject traffic into the network. While the goal of these platforms is partially overlapped with that of JingZhao, we are aiming at full-system integration, especially the interaction between system software and the NIC. Hence we provide both system-level simulation and provide standard interfaces to software. Other platforms are based on SmartNICs, which provide software interfaces to talk to the NIC. However, as discussed in Section 2.3.2, due to the weakness of the on-NIC ARM cores, these platforms provide more flexibility rather than performance.

The work most similar to ours is Corundum[36], which is a famous open-sourced Ethernet NIC. Corundum focuses on the stateless function of packet processing, including checksum offloading, receive-side-scaling, with little concern on stateful functions. Besides, the components of Corundum are tightly coupled and specific for Ethernet processing, which are hard to be reused. Inspired by Corundum, we step a little further to generalize several reusable building blocks and provide a complete system-level framework. Another similar work is SRNIC[16], which implements a RDMA NIC as we do, however, SRNIC focuses on solving the scalability problem of RDMA, not NIC prototyping. Besides, SRNIC is a commercial product, JingZhao is completely open-sourced.

### 7.2 Domain-Specific Networks

The distance between Computing Unit and Communication Unit is shortened with new architectures and protocols developed, such Nvidia NVLink[39] and Intel Gaudi AI Processor[40]. The NIC is tended to be tightly coupled with the XPU (X stands for domain-specific application). From the perspective of XPU, there are two types of semantics to exchange data with a remote peer: Load/Store for scale-up networks, and Remote Read/Write for scale-out networks. JingZhao provides a NIC reference design used in scale-out networks, and could be further modified to support scale-up networks.

### 7.3 Network Simulation

Traditional network simulators such as OMNeT++[22] and NS3[23], focus on high-level behavior of the protocols, but can hardly reflect how the NICs interact with the host CPU and memory. Our framework is a complement to these works, which helps researchers figure out how network traffic patterns affect the architectural design micro-architecture of the NIC and how a dedicated on-NIC cache hierarchy affects network performance.

## 8 Conclusion and Future Work

The network is becoming Domain-Specific, which requires on-demand design of the network protocols, as well as the microarchitecture of the NICs. To accelerate the process of designing, verifying and optimizing a domain-specific NIC, we have proposed JingZhao, an open-sourced framework for NIC prototyping, which could be leveraged to rapidly implement a domain-specific NIC.

JingZhao provides several reusable, building blocks, as well as a full-fledged RDMA NIC, to help rapidly prototype a high-performance NIC. Our evaluation results show that new network functions can be easily integrated into the framework, and achieve line-rate packet processing. We expect JingZhao to be a fundamental tool to accelerate the deployment of innovative network protocols.

In the future, we will continue explore the design space of NIC prototyping in three directions:

1) Add more support for protocol-specific building blocks, which could become a fundamental library for designing a high-performance NIC, just as “Glibc” in software development;

2) Due to our non-blocking resource access mechanism, the bandwidth of JingZhao NIC is stable even there’s heavy

cache miss, however, throughput and latency still needs to be improved. A possible solution is to optimize the algorithm of cache replacement, which may reduce cache miss rate;

3) JingZhao is now aimed at system-level interconnection, which is also called "Scale-Out" network. We are considering adding support for load/store instructions, which could be used to interconnect GPUs or NPUs in a "Scale-Up" network.

**References**

[1] Bojie Li. High Performance Data Center Systems with Programmable Network Interface Cards. https://01.me/files/pubs/bojieli-phd-thesis-en-auto-translated.pdf, Accessed 2024-10-01

[2] Gangidi A, Miao R, Zheng S, et al. RDMA over Ethernet for Distributed Training at Meta Scale[C]//Proceedings of the ACM SIGCOMM 2024 Conference. 2024: 57-70.

[3] Jiang Z, Lin H, Zhong Y, et al. MegaScale: Scaling large language model training to more than 10,000 GPUs[C]//21st USENIX Symposium on Networked Systems Design and Implementation. 2024: 745-760.

[4] Qian K, Xi Y, Cao J, et al. Alibaba HPN: a data center network for large language model training[C]//Proceedings of the ACM SIGCOMM 2024 Conference. 2024: 691-706.

[5] Kalia A, Kaminsky M, Andersen D G. Design guidelines for high performance RDMA systems[C]//2016 USENIX Annual Technical Conference. 2016: 437-450.

[6] Arslan S, Li Y, Kumar G, et al. Bolt: Sub-RTT Congestion Control for Ultra-Low Latency[C]//20th USENIX Symposium on Networked Systems Design and Implementation. 2023: 219-236.

[7] Zhang Y, Meng Q, Hu C, et al. Revisiting Congestion Control for Lossless Ethernet[C]//21st USENIX Symposium on Networked Systems Design and Implementation. 2024: 131-148.

[8] Xu J, Wang Z, Yang F, et al. FNCC: Fast Notification Congestion Control in Data Center Networks[C]//Proceedings of the 53rd International Conference on Parallel Processing. 2024: 127-137.

[9] Xia J, Zeng G, Zhang J, et al. Rethinking transport layer design for distributed machine learning[C]//Proceedings of the 3rd Asia-Pacific Workshop on Networking. 2019: 22-28.

[10] Li J, Gong H, De Marchi F, et al. Uniform-Cost Multi-Path Routing for Reconfigurable Data Center Networks[C]//Proceedings of the ACM SIGCOMM 2024 Conference. 2024: 433-448.

[11] Hu Q, Ye Z, Wang Z, et al. Characterization of large language model development in the datacenter[C]// Proceedings of the 21st USENIX Symposium on Networked Systems Design and Implementation. 2024: 709-729.

[12] Joshi R, Song C H, Khooi X Z, et al. Masking corruption packet losses in datacenter networks with link-local retransmission[C]//Proceedings of the ACM SIGCOMM 2023 Conference. 2023: 288-304.

[13] Mittal R, Shpiner A, Panda A, et al. Revisiting network support for RDMA[C]//Proceedings of the 2018 Conference of the ACM Special Interest Group on Data Communication. 2018: 313-326.

[14] Lu Y, Chen G, Ruan Z, et al. Memory efficient loss recovery for hardware-based transport in datacenter[C]//Proceedings of the First Asia-Pacific Workshop on Networking. 2017: 22-28.

[15] Ma J, Guo Z, Pan Y, et al. ORNIC: A High-Performance RDMA NIC with Out-of-Order Packet Direct Write Method for Multipath Transmission[J]. Electronics (2079-9292), 2025, 14(1).

[16] Wang Z, Luo L, Ning Q, et al. SRNIC: A scalable architecture for RDMA NICs[C]//20th USENIX Symposium on Networked Systems Design and Implementation. 2023: 1-14.

[17] Chen Y, Lu Y, Shu J. Scalable RDMA RPC on reliable connection with efficient resource sharing[C]//Proceedings of the Fourteenth EuroSys Conference 2019. 2019: 1-14.

[18] Kang N, Wang Z, Yang F, et al. csRNA: connection-scalable RDMA NIC architecture in datacenter environment[C]//2022 IEEE 40th International Conference on Computer Design. 2022: 398-406.

[19] Pan P, Chen G, Wang X, et al. Towards stateless rnic for data center networks[C]//Proceedings of the 3rd Asia-Pacific Workshop on Networking. 2019: 57-63.

[20] Kong X, Zhu Y, Zhou H, et al. Collie: Finding performance anomalies in RDMA subsystems[C]//19th USENIX Symposium on Networked Systems Design and Implementation. 2022: 287-305.

[21] Liu K, Jiang Z, Zhang J, et al. Hostping: Diagnosing intra-host network bottlenecks in RDMA servers[C]//20th USENIX Symposium on Networked Systems Design and Implementation. 2023: 15-29.

[22] Varga A. OMNeT++[M]//Modeling and tools for network simulation. Berlin, Heidelberg: Springer Berlin Heidelberg, 2010: 35-59.

[23] Kumar A R A, Rao S V, Goswami D. NS3 simulator for a study of data center networks[C]//2013 IEEE 12th International Symposium on Parallel and Distributed Computing. IEEE, 2013: 224-231.

[24] Bi H, Wang Z H. Dpdk-based improvement of packet forwarding[C]//ITM web of Conferences. EDP Sciences, 2016, 7: 01009.

[25] Rizzo L. netmap: a novel framework for fast packet I/O[C]//21st USENIX Security Symposium (USENIX Security

12). 2012: 101-112.
[26] Qiu Y, Xing J, Hsu K F, et al. Automated smartnic offloading insights for network functions[C]//Proceedings of the ACM SIGOPS 28th Symposium on Operating Systems Principles. 2021: 772-787.
[27] Wei X, Cheng R, Yang Y, et al. Characterizing Off-path SmartNIC for Accelerating Distributed Systems[C]//17th USENIX Symposium on Operating Systems Design and Implementation. 2023: 987-1004.
[28] Lin W, Shan Y, Kosta R, et al. SuperNIC: An FPGA-Based, CloudOriented SmartNIC[C]//Proceedings of the 2024 ACM/SIGDA International Symposium on Field Programmable Gate Arrays. 2024: 130-141.
[29] Forencich A, Snoeren A C, Porter G, et al. Corundum: An open-source 100-gbps nic[C]//2020 IEEE 28th Annual International Symposium on Field-Programmable Custom Computing Machines. IEEE, 2020: 38-46.
[30] AMD, "AlveoLink," https://github.com/Xilinx/AlveoLink, Accessed: 2023-11-01.
[31] AMD, "XUP Vitis Network Example," https://github.com/Xilinx/xup vitis network example, Accessed: 2023-11-01.
[32] D. Korolija, T. Roscoe and G. Alonso, "Do OS abstractions make sense on FPGAs?," 14th USENIX Symposium on Operating Systems Design and Implementation (OSDI 20), 2020, pp. 991–1010
[33] Z. Wang, H. Huang, J. Zhang, F. Wu and G. Alonso, "FpgaNIC: An FPGA-based Versatile 100Gb SmartNIC for GPUs," 2022 USENIX Annual Technical Conference, Carlsbad, CA, 2022, pp. 967-986
[34] V. Krishnan, O. Serres and M. Blocksome, "COnfigurable Network Protocol Accelerator (COPA): An Integrated Networking/Accelerator Hardware/Software Framework," 2020 IEEE Symposium on High Performance Interconnects (HOTI), Piscataway, NJ, USA, 2020, pp. 17-24
[35] AMD, "qep-driver," https://github.com/Xilinx/qep-drivers, Accessed: 2023-11-01.
[36] Forencich A, Snoeren A C, Porter G, et al. Corundum: An open-source 100-gbps nic[C]//2020 IEEE 28th Annual International Symposium on Field-Programmable Custom Computing Machines. IEEE, 2020: 38-46.
[37] Ruiz M, Sidler D, Sutter G, et al. Limago: An FPGA-based open-source 100 GbE TCP/IP stack[C]//2019 29th International Conference on Field Programmable Logic and Applications. IEEE, 2019: 286-292.
[38] Zhong G, Kolekar A, Amornpaisannon B, et al. A primer on reconic: Rdma-enabled compute offloading on smartnic[J]. arXiv preprint arXiv:2312.06207, 2023.
[39] NVIDIA. Nvlink and nvlink switch [Z]. 2024.
[40] Chen Levkovich. The habana gaudi2 processor for deep learning [Z]. 2024.
[41] Li B, Ruan Z, Xiao W, et al. Kv-direct: High-performance in-memory key-value store with programmable nic[C]//Proceedings of the 26th Symposium on Operating Systems Principles. 2017: 137-152.
[42] Yang F, Wang Z, Ma X, et al. SwitchAgg: A further step towards in-network computing[C]//2019 IEEE Intl Conf on Parallel & Distributed Processing with Applications, Big Data & Cloud Computing, Sustainable Computing & Communications, Social Computing & Networking. IEEE, 2019: 36-45.
[43] Linux-Rdma. 2024. GitHub - linux-rdma/perftest: Infiniband Verbs Performance Tests. https://github.com/linuxrdma/perftest

## Appendix A: Discussion of fundamental primitives

Modern NIC pipelines are typically composed of multiple hardware modules connected in series or parallel. These modules could be further classified as protocol-dependent and protocol-independent. To accelerate NIC pipeline design and implementation, we require higher-level design primitives or templates that are tailored for network transmission functions while remaining protocol-independent.

Existing IP block abstractions operate at a level which lacks available abstraction for network processing logic, which still necessitates significant development effort. We observe that most available IPs are either proprietary or exhibit substantial functional gaps when applied to packet processing requirements. Taking RDMA NICs as an example, packet processing involves extensive data scatter-gather operations with host memory, requiring not only PCIe transaction handling but also frequent data segmentation and alignment - operations particularly prone to edge case oversights during implementation. Furthermore, RDMA's connection-oriented operations demand on-chip management of numerous connection states, logically requiring a dynamic multi-queue structure analogous to software-linked lists. However, current IPs only provide single-FIFO interfaces, mandating additional development effort. The abstracted hardware primitives are protocol-independent, and could be configured through parameterization to adapt to diverse processing needs. Below demonstrates a hardware template for dynamic multi-queues, where users can easily customize queue quantity, depth, and bit-width to accommodate dynamic connection state storage requirements.

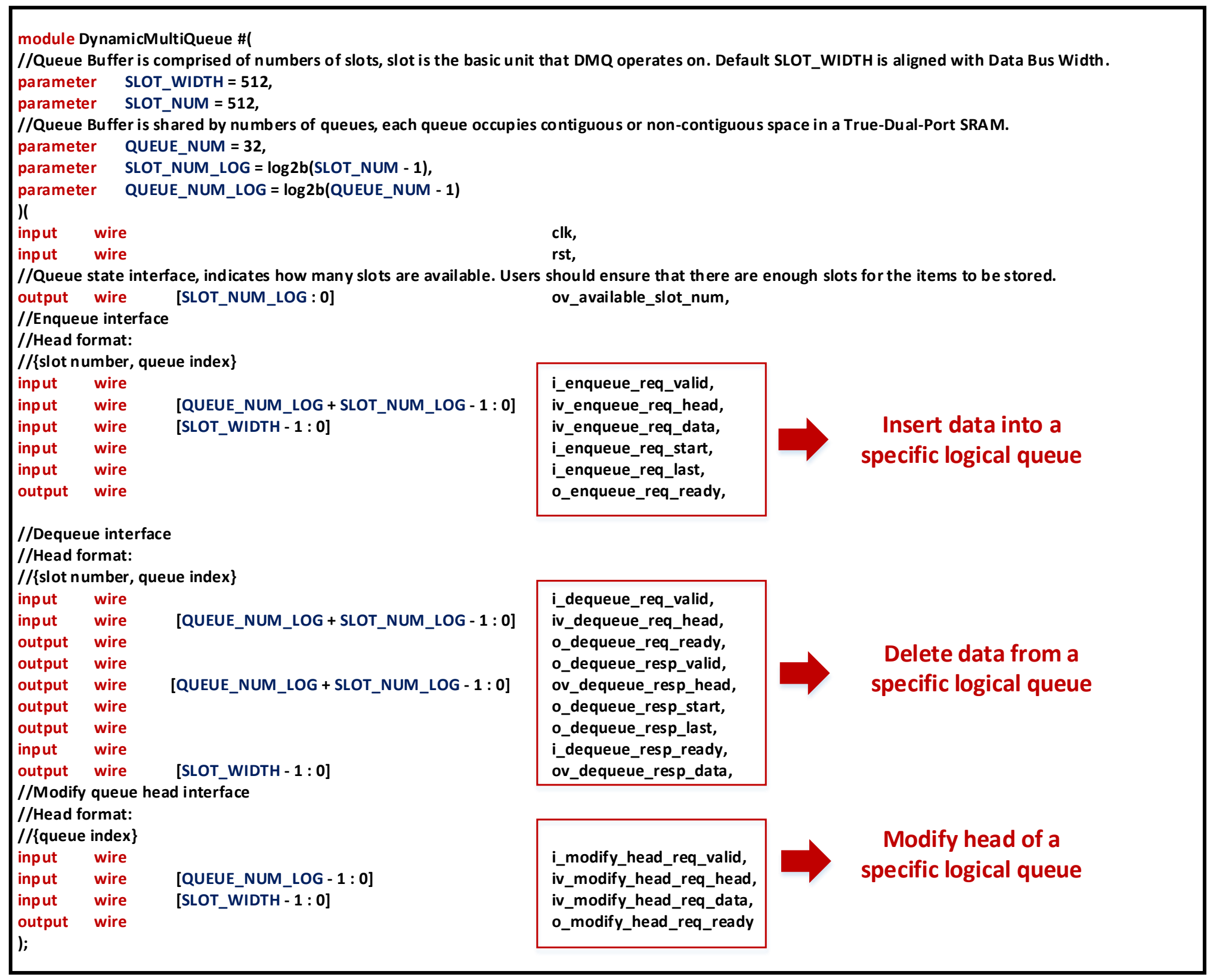

**Figure 12: A Hardware Template for Dynamic Multi-Queue**

## Appendix B: Memory layout, user interfaces and parameters of SHOST

In SHOST, each host independently maintains a linear address space for dynamically creating and releasing Data Buffers, Queue Buffers, and ICM Buffers. The host establishes address mappings to simulate the operating system's virtual-to-physical address translation. Currently, we implement a simplified mapping mechanism rather than complex address translation - virtual and physical addresses maintain a fixed linear offset relationship. While basic, this approach effectively validates network protocols that utilize virtual address spaces for memory access. Taking ICM space as an example, each host's ICM address range follows the allocation shown in the table below. Users can dynamically repartition the space sizes according to specific requirements.

**Table 5: Memoy Layout of SHOST**

| Offset in ICM Space | Contents |
|---|---|
| 64’h0000_0000_0000_0000 | Memory Translation Table |

| 64’h0000_1000_0000_0000 | QP Context |
|---|---|
| 64’h0000_2000_0000_0000 | CQ Context |
| 64’h0000_3000_0000_0000 | Memory Protection Table |

Researchers could use the combination of different parameters to generate test cases. The parameters are listed below.

**Table 6: Parameters of Test Cases**

| Parameters | Description | Range |
|---|---|---|
| +HCA_CASE_NAME | Name of the test case. | String |
| +HCA_HOST_NUM | Number of hosts instantiated by SHOST. | [1, 256](1 for loopback test) |
| +HCA_PROC_NUM | Number of processes on each host. | [2-2048] |
| +HCA_DB_NUM | Number of doorbells of each QP. | [1, 1000] |
| +HCA_PAGE_NUM | Number of memory pages of each communication request. | [1-1000] |
| +RC_QP_NUM | Number of Reliable Connections. | [0, 16384] |
| +UC_QP_NUM | Number of Unreliable Datagrams. | [0, 16384] |
| +UD_QP_NUM | Number of Unreliable Connections. | [0,16384] |
| +WRITE_WQE_NUM | Number of RDMA Write requests. | [0, 1024] |
| +READ_WQE_NUM | Number of RDMA Read requests. | [0, 1024] |
| +SEND_WQE_NUM | Number of Send requests. | [0, 1024] |
| +RECV_WQE_NUM | Number of Recv requests. | [0, 1024] |

**Appendix C: System Architecture of JingZhao FPGA Prototype**

The System Architecture of JingZhao FPGA Prototype is shown below. A kernel-space driver (integrated into Linux InfiniBand subsystem) handles QP/CQ creation, memory allocation/registration and hardware initialization (control path), and a user-level driver (hooked to libibverbs) directly talk to the NIC and implements the Post_Send/Post_Recv/Poll_CQ functions (data path). They work collaboratively to be fully compatible to standard OFED verbs.

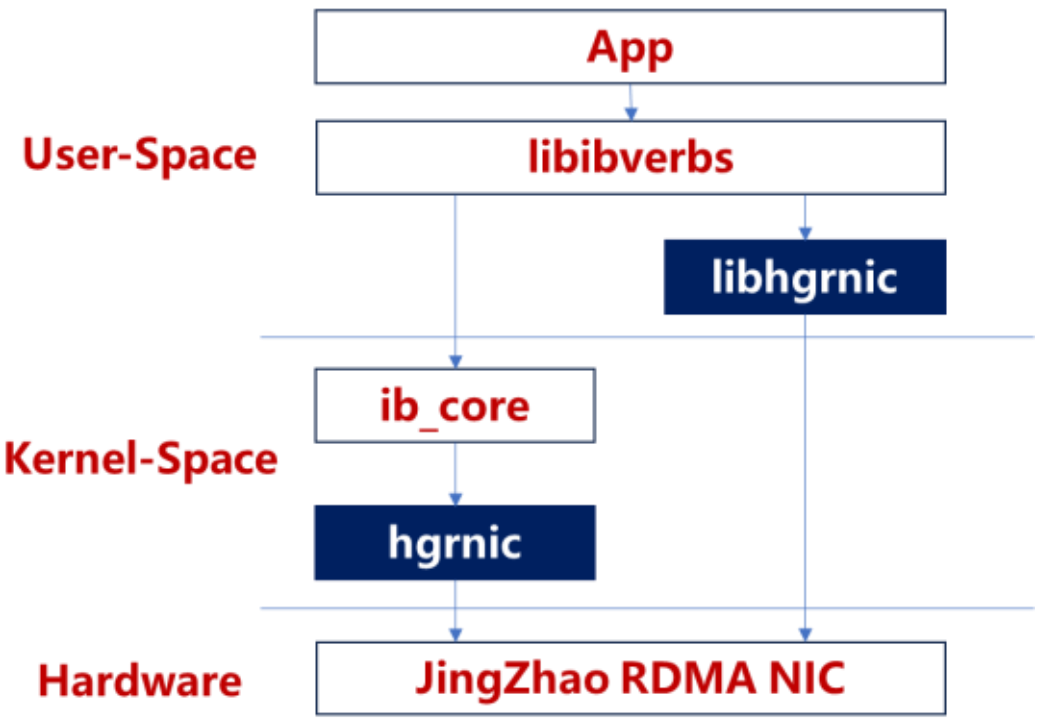

**Figure 13: System Architecture for JingZhao FPGA Prototype**

**Appendix D: Silicon Validation of JingZhao NIC**

Leveraging the building blocks, and verification framework provided by JingZhao, we have successfully developed a RDMA NIC compatible with standard OFED libibverbs APIs. After FPGA prototyping, we use EDA tools to generate netlists and GDS files of our NIC, which achieved timing closure at 450MHz. Furthermore, the design was taped out using TSMC's 28nm process, yielding a chip with a total area of 43mm². Thanks to the reusable components and primitives in JingZhao's design framework, we accomplished the implementation of RDMA functionality (developed by 3 Ph.D candidates) in less than three months.

In post-silicon validation, we use the Perftest benchmarking tool to measure the latency and bandwidth of the chip. The results demonstrate a peak performance of (27.5Gbps, 2.5μs) for RDMA Write and （26.8Gbps, 2.7μs） for RDMA Read. To our regret, due to PCB (Printed Circuit Board) design problems, the PCIe interface of our NIC currently operates stably at Gen3 x4 (designed for Gen4 x 8), resulting in a theoretical maximum bandwidth of 32Gbps, the rectification of this problem is in progress. The figure below shows (from left to right) the GDS layout file and substrate design diagram.

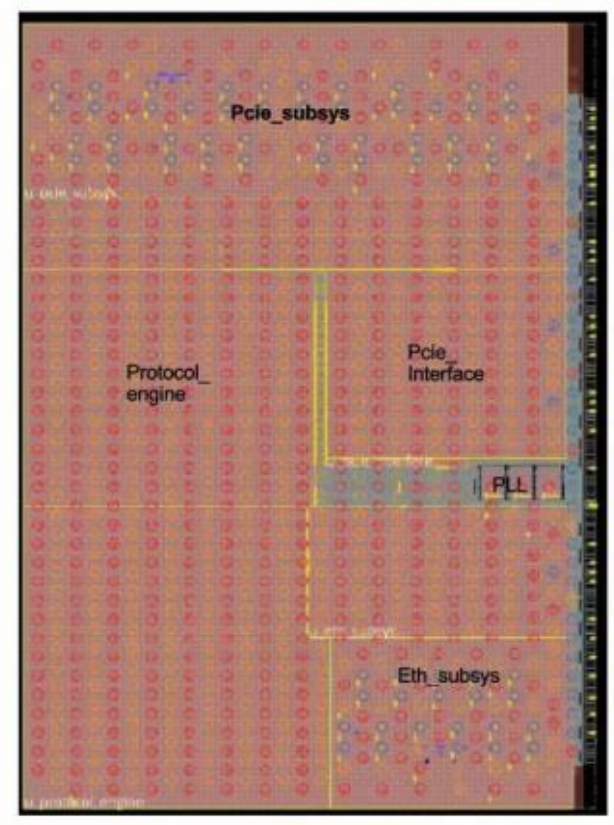

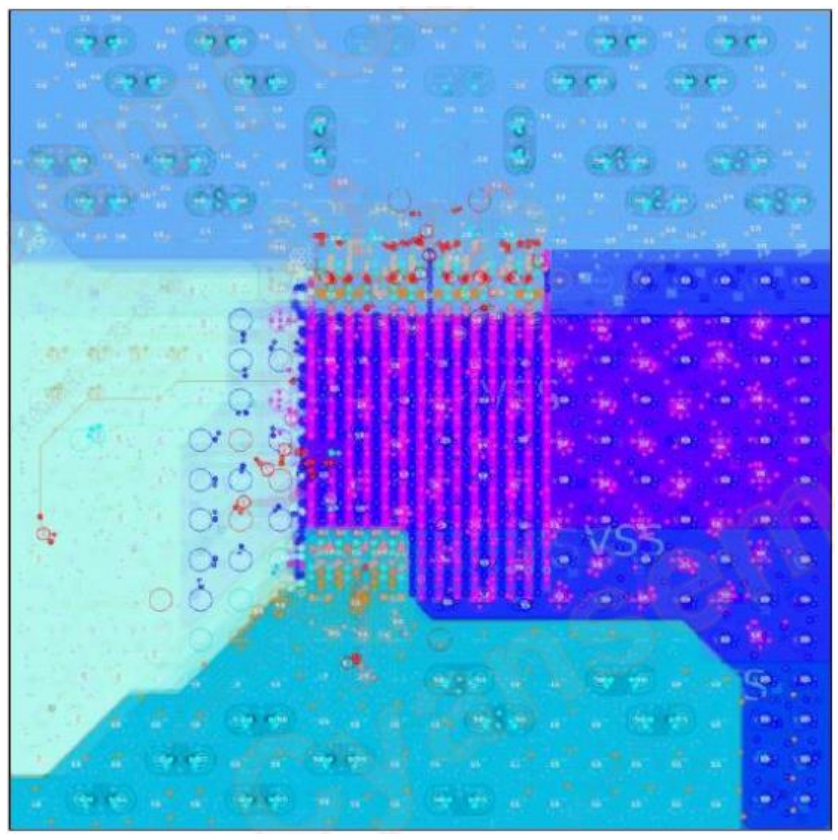

**Figure 14: GDS Layout and Substrate Design Diagram**

**Appendix E: Discussion of RecoNIC, Corundum and JingZhao**

RecoNIC is a FPGA-based in-network computing acceleration framework developed by AMD, capable of offloading application logic to the NIC for execution. RecoNIC aims to provide researchers with a development and validation framework for high-performance NICs. Specifically, targeting the three objectives discussed in Section 2, there are several key differences between JingZhao and RecoNIC:

1) High-performance protocol offloading. RecoNIC extends RDMA communication by supporting computational logic offloading. To achieve this, it integrates an RDMA engine into the NIC and incorporates both a Streaming Compute and a Lookaside Compute accelerator for task offloading. However, its network protocol processing functions (such as queue scheduling and communication semantics) are tightly coupled with RDMA and cannot be modified. In contrast, JingZhao provides fundamental network processing primitives, allowing users to construct custom transport protocols. Additionally, JingZhao optimizes memory access operations for stateful protocol processing, implementing an efficient on-chip cache-to-host memory interaction architecture.

2) Agile Development. As an industry product, RecoNIC supports multiple programming languages including P4, HLS, and RTL, offering high programming flexibility for computational task offloading. However, it lacks built-in support for fundamental network functional components, requiring users to implement each subsystem functionality themselves. In contrast, JingZhao abstracts protocol-independent processing primitives and subsystems, primarily targeting network protocol offloading. Researchers only need minimal RTL development and subsystem configuration to construct an efficient processing pipeline.

3) System-level Simulation. RecoNIC only provides FPGA-based prototype verification, which presents significant debugging challenges. In contrast, JingZhao offers not only a FPGA verification platform, but a complete system simulation framework. This framework can fully replicate the interaction process between drivers, communication libraries, and NIC hardware through the system bus, enabling easier performance evaluation.

Moreover, RecoNIC is only partially open-sourced (its RDMA module remains proprietary) and is not compatible to OFED libfabrics, whereas JingZhao's entire software and hardware framework is fully open-sourced can compatible to standard verbs API.

Corundum is a another renowned open-source NIC framework that implements a standard Ethernet card on multiple FPGA platforms with Socket interface compatibility. Designed as an FPGA-based platform to assist researchers in NIC architecture exploration, Corundum serves as an excellent reference implementation. Building upon Corundum, numerous novel network protocols and architectures have been developed. However, developing or porting new protocols to Corundum presents significant challenges due to three primary limitations: 1) It's fixed pipeline architecture contains components specifically optimized for TCP/IP packet processing, offering limited flexibility for other protocols; 2) The framework lacks support for stateful protocols, substantially restricting its application scope; 3) While providing a Python-based simulation environment, Corundum directly injects network traffic into the NIC model without accurately simulating system software interactions. Similar to RecoNIC, system-level validation must be performed on FPGA prototypes, which introduces significant debugging challenges.

JingZhao systematically addresses these limitations through three key innovations: 1) Abstracting versatile processing primitives into reusable components that enable pipeline construction via simple parameter configuration and interconnection; 2) Implementing stateful protocol support through non-blocking memory access structures that facilitate high-performance on-chip caching and host memory interactions; 3) Accurately modeling PCIe-based interactions between the NIC, host memory, and drivers, enabling more accurate performance evaluation and pre-FPGA functional/performance debugging. While JingZhao's design draws substantial inspiration from Corundum - having incorporated many of its architectural concepts - we expressly acknowledge our intellectual debt to the Corundum authors and extend sincere gratitude for their foundational contributions.

## Appendix F: FPGA Resource Utilization of JingZhao Prototype

**Table 7: FPGA Resource Usage (VU37P)**

| | Components | | Configuration (Buffer width * depth) | LUT (%) | FFs (%) | BRAM (%) |
|---|---|---|---|---|---|---|
| Primitive | Append Header | | / | 651.84 (0.05%) | 1303.68 (0.05%) | 0 |
| | Remove Header | | / | 260.736 (0.02%) | 782.208 (0.03%) | 0 |
| | Scatter Data | | 512 * 128, buffering data stream | 8473.92 (0.65%) | 2085.888 (0.08%) | 4.512 (0.47%) |
| | Gather Data | | 512 * 128, buffering data stream | 6518.4 (0.5%) | 2085.888 (0.08%) | 5.472 (0.57%) |
| | Dynamic Multi-Queue | | 512 * 32, shared queue buffer | 15774.528 (1.21%) | 2346.624 (0.09%) | 22.56 (2.35%) |
| | Dynamic Buffer | | 512 * 32, shared data buffer | 17208.576 (1.32%) | 1042.944 (0.04%) | 22.56 (2.35%) |
| Subsystem | DMA | | 512 * 512, reorder buffer for Cpl | 69095.04 (5.3%) | 75352.704 (2.89%) | 60.48 (6.3%) |
| | Queue Subsystem | | 128 * 1024, queue cache | 17338.944 (1.33%) | 7300.608 (0.28%) | 18.048 (1.88%) |
| | Transport Subsystem | Go-Back-N | 512 * 32, egress buffer | 4171.776 (0.32%) | 9125.76 (0.35%) | 1.44 (0.15%) |
| | | Selective Repeat | 512 * 32, ingress/egress buffer | 2346.624 (0.18%)) | 10690.176 (0.41%) | 2.88 (0.3%) |
| | Resource Subsystem | QPC Memory | 416 * 128, QPC cache buffer | 4432.512 (0.34%) | 3650.304 (0.14%) | 20.448 (2.13%) |
| | | CQC Memory | 128 * 128, CQC cache buffer | 1955.52 (0.15%) | 1564.416 (0.06%) | 8.832 (0.92%) |
| | | MPT Memory | 256 * 512, MPT cache buffer | 2216.256 (0.17%) | 1825.152 (0.07%) | 15.936 (1.66%) |
| | | MTT Memory | 64 * 1024, MTT cache buffer | 1825.152 (0.14%) | 1564.416 (0.06%) | 18.336 (1.91%) |
| | | Value Memory | 256 * 1024, Value cache buffer | 3650.304 (0.28%) | 3128.832 (0.12%) | 18.72 (1.96%) |
| Example Design | RDMA Core | | / | 46280.64 (3.55%) | 18512.256 (0.71%) | 39.264 (4.09%) |
| | Key-Value Core | | / | 34808.256 (2.67%) | 33374.208 (1.28%) | 31.2 (3.25%) |